\documentclass[aps,prb,twocolumn,shortbibliography,superscriptaddress]{revtex4-1}
\usepackage{epsfig}
\usepackage{epstopdf}
\usepackage{amsmath}
\usepackage{amsfonts}
\usepackage{amssymb}
\usepackage{hyperref} 
\usepackage{bm}
\usepackage{makecell}
\usepackage{rotating}
\usepackage{hyperref}
\usepackage{multirow}
\usepackage{graphicx}
\usepackage{dcolumn}
\usepackage{bm}
\usepackage{color}

\usepackage{tikz,xcolor,hyperref}

\definecolor{lime}{HTML}{A6CE39}
\DeclareRobustCommand{\orcidicon}{%
	\begin{tikzpicture}
	\draw[lime, fill=lime] (0,0)
	circle [radius=0.16]
	node[white] {{\fontfamily{qag}\selectfont \tiny ID}};
	\draw[white, fill=white] (-0.0625,0.095)
	circle [radius=0.007];
	\end{tikzpicture}
	\hspace{-2mm}
}

\foreach \x in {A, ..., Z}{%
	\expandafter\xdef\csname orcid\x\endcsname{\noexpand\href{https://orcid.org/\csname orcidauthor\x\endcsname}{\noexpand\orcidicon}}
}

\begin{document}

\title{Interplay between Relativistic Spin-Momentum Locking and Breaking \\ of Inversion Symmetry: conditions for p-wave magnetism}

\author{Amar Fakhredine\orcidF}
\affiliation{Institute of Physics, Polish Academy of Sciences, Aleja Lotnik\'ow 32/46, 02668 Warsaw, Poland}

\author{Giuseppe Cuono\orcidA}
\affiliation{Consiglio Nazionale delle Ricerche (CNR-SPIN), Unit\'a di Ricerca presso Terzi c/o Universit\'a “G. D’Annunzio”, 66100 Chieti, Italy}

\author{Jan Skolimowski\orcidG}
\affiliation{International Research Centre Magtop, Institute of Physics, Polish Academy of Sciences,
Aleja Lotnik\'ow 32/46, PL-02668 Warsaw, Poland}

\author{Silvia Picozzi\orcidP}
\affiliation{Department of Materials Science, University Milan-Bicocca, 20125 Milan (IT)}
\affiliation{Consiglio Nazionale delle Ricerche (CNR-SPIN), Unit\'a di Ricerca presso Terzi c/o Universit\'a “G. D’Annunzio”, 66100 Chieti, Italy}


\author{Carmine Autieri\orcidD}
\email{autieri@magtop.ifpan.edu.pl}
\affiliation{International Research Centre Magtop, Institute of Physics, Polish Academy of Sciences,
Aleja Lotnik\'ow 32/46, PL-02668 Warsaw, Poland}
\affiliation{SPIN-CNR, UOS Salerno, IT-84084 Fisciano (SA), Italy}

\date{\today}
\begin{abstract}
We investigate the interplay between relativistic spin-momentum locking arising from altermagnetism and various forms of inversion symmetry breaking. Depending on the symmetry breaking, this can give rise to Rashba-type spin-orbit coupling (SOC), Weyl-type SOC, or the coexistence of two distinct spin-momentum lockings. We focus on the altermagnetic Ca$_2$RuO$_4$ as a testbed material. Our results reproduce the experimentally observed ground state, which is an A-centered magnetic order with the Néel vector along the $b$-axis, hosting spin cantings along the $a$- and $c$-axes but without weak ferromagnetism. Ca$_2$RuO$_4$ exhibits relativistic spin-momentum locking, with different even-parity wave orders for the three spin components. We interpret the experimental results on doped samples as evidence for a transition from a pure altermagnetic phase to a weak ferromagnetic phase.
Under ferroelectric- and antiferroelectric-like distortions, there are no qualitative changes in the non-relativistic spin-momentum locking and in the weak ferromagnetism. However, we observe the rise of the Rashba or Weyl-type SOC. Using numerical and analytical models, we investigate which nodal planes persist when inversion symmetry is broken in the relativistic case. The spin-momentum locking of the other components adopt a p-wave character in the case of Rashba; in contrast, Weyl-type SOC disrupts all nodal planes, leaving only nodal lines.
Finally, to simulate a stripe phase with structural distortions along the $z$-axis, we studied a modulated electric field inducing atomic displacements within one Ca$_2$RuO$_4$ layer. This produces a magnetic phase transition to an exotic altermagnetic state with two non-relativistic spin-momentum lockings hosting weak ferromagnetism. Our research presents a comprehensive analysis of various possible scenarios in altermagnets with breaking of inversion symmetries under relativistic effects.
\end{abstract}

\pacs{}

\maketitle

\section{Introduction}

The comprehension and classification of the non-relativistic spin-momentum locking in altermagnets is one of the most striking achievements in the field of magnetism in recent years\cite{Smejkal22,hayami2019momentum,yuan2023degeneracy}. 
The study of the altermagnetism was extended to the anomalous Hall conductivity\cite{PhysRevLett.130.036702,reichlova2024observation}, weak ferromagnetism\cite{PhysRevLett.132.176702,839n-rckn}, spin Hall conductivity\cite{liao2024separation}, spin-current generation\cite{PhysRevLett.126.127701}, spin-to-charge conversion\cite{bai2023efficient}, surfaces\cite{D3NR03681B}, low dimensionality\cite{xu2025chemicaldesignmonolayeraltermagnets,PhysRevB.111.184407}, interfaces\cite{doi:10.1021/acs.nanolett.4c02248} and topology\cite{PhysRevLett.134.096703,Li2025}.
A variety of devices for spintronic applications have been proposed, exploiting altermagnetic materials and other systems which break time-reversal symmetry while retaining a zero net magnetization\cite{PhysRevX.12.011028,jungwirth2025altermagneticspintronics,Song2025,doi:10.1021/jacs.4c14503,Berritta2025}.
Recently, several authors have addressed the evolution of the non-relativistic spin-momentum locking in the relativistic case\cite{AutieriRSML,Fernandes2024,PhysRevB.110.144412}, where a powerful tool is the multipole analysis\cite{hirakida2025multipoleanalysisspincurrents,Hayami2024}. Therefore, we define the relativistic spin-momentum locking as the combination of the spin-momentum locking for the S$_x$, S$_y$ and S$_z$ components.  

Another important aspect investigated in the field of altermagnetism is the effect of breaking inversion symmetry by interfaces, spontaneous ferroelectricity, or an external electric field.
Recent works have been focusing on altermagnetism in polar systems\cite{kim2023observation}, electric-field induced altermagnetism in 2D systems \cite{PhysRevB.108.L180403,mazin2023induced,D3NR03681B}, and in 3D systems\cite{D4TC00899E}. Studies were performed on ferroelectrics \cite{smejkal2024altermagneticmultiferroicsaltermagnetoelectriceffect,D4MH01619J} and antiferroelectrics\cite{PhysRevLett.134.106801}. Manipulation of the weak ferromagnetism by applied electric field was widely investigated in the past\cite{PhysRevLett.100.167203}. In the recent literature, weak ferromagnetism was revisited, incorporating insights from research on altermagnetism.\cite{PhysRevB.107.075202}
It was demonstrated that even tiny shifts of the atoms with a consequent breaking of the crystal symmetries can enhance the anomalous Hall effect in a non-collinear antiferromagnet \cite{https://doi.org/10.1002/adma.202209616}, and that an electric field can generate a magnetic toroidal moment in altermagnets\cite{hayashida2024electricfieldinduced}. A recent study has proposed an all-electrical manipulation of spin-polarized currents in altermagnetic bilayers, enabling gate-tunable and reversible spin-polarization under an out-of-plane electric field, allowing for a new path towards altermagnetic-based spintronic devices\cite{peng2025all}. Yet another study shows that 2D altermagnets exhibit anisotropic spin-transport where the application of strain or an electric field induces conductivity in spin-up and spin-down channels along orthogonal directions. This enables highly controllable spin currents in 2D altermagnetic devices\cite{zhang2025tunable}.

Previous studies of altermagnetic systems without inversion symmetry have investigated the presence of the spin-momentum locking\cite{D3NR04798A}, magnetic phase transition\cite{leon2025strainenhancedaltermagnetismca3ru2o7}, the interplay between altermagnetism and a persistent spin helix\cite{tenzin2025persistentspintexturesaltermagnetism}, weak ferromagnetism\cite{jo2025weak}, or ferroelectricity\cite{smejkal2024altermagneticmultiferroicsaltermagnetoelectriceffect,https://doi.org/10.1002/adfm.202505813,D4MH01619J}.
Among the most exotic effects arising from the breaking of inversion symmetry, there are the Rashba\cite{Picozzi:2014_FP,PhysRevLett.115.256801,D3CP04242A} and Weyl effect\cite{Roy2022,wu2025unconventional}.
Rashba altermagnets were studied, including only the non-relativistic spin-momentum locking using model Hamiltonians\cite{chen2024helicitycontrolledspinhall,mukherjee2025electricfieldcontrolledsecondorder}.
What is missing in the literature is a thorough investigation of the interplay between altermagnetic spin-splitting and Rashba-, or Weyl-type spin-splitting.
To fill this gap, we begin by considering the centrosymmetric altermagnet Ca$_2$RuO$_4$ as a testbed material for the study of the interplay between relativistic spin-momentum locking and the breaking of inversion symmetry. Indeed, this material exhibits relativistic spin-momentum locking and hosts several internal degrees of freedom (also due to the presence of four magnetic atoms), offering the possibility of having different kinds of inversion symmetry breaking, even if some are experimentally unlikely. We aim to analyze how different types of inversion-symmetry breaking give rise to the interplay between relativistic spin-momentum locking, Rashba, and Weyl-type relativistic spin-splitting.

The quasi-two-dimensional Ca$_2$RuO$_4$ is one of the few altermagnets with 4d electrons\cite{Cuono23orbital}. It is an orbital-selective altermagnet in which the ${xy}$ bands are not altermagnetic to the first order, while the $\gamma$z ($\gamma$ = x,y) bands display altermagnetism \cite{Cuono23orbital,Cuono25}. Ca$_2$RuO$_4$ allows two magnetic configurations with zero net magnetization, referred to as A-centered and B-centered. The experimentally observed ground state corresponds to the A-centered order, characterized by magnetic moments aligned along the $b$-axis, a small canting along $c$, and no canting along $a$. \cite{Braden98,Porter18}.
Although the magnetic ground state of the bulk (space group 61) shows orbital-selective altermagnetism with no altermagnetism in the $xy$ orbitals\cite{Cuono23orbital}, the single layer (space group 7) shows altermagnetism in these orbitals\cite{D4NR04053H}. The altermagnetic phase switches from planar d$_{xz}$-wave in the bulk\cite{Cuono23orbital} to planar d$_{xy}$-wave in the two-dimensional case\cite{D4NR04053H}.
To avoid confusion, in the paper, we will use the notation d$_{xy}$, d$_{xz}$ and d$_{yz}$ for the spin-momentum locking, while $xy$, $xz$ and $yz$ will be used for the bands.
Beyond altermagnetism, Ca$_2$RuO$_4$ exhibits other exotic properties, such as a metal-insulator transition at 357 K \cite{ricco2018situ,Alexander99} and non-linear responses to current and electric field. The Mott transition can also occur at lower temperatures under an external electric field or currents \cite{Nakamura13,Zhang19,Mattoni20,Cirillo19,Cuoco06,Forte10,Cuono22,doi:10.1021/acsami.0c05181}. The metal-insulator transition is accompanied by a structural transition from a phase with a short $c$-axis to a phase with a long $c$-axis, both belonging to the same orthorhombic space group no. 61\cite{Gorelov10,Zhang17,Porter18,Okazaki13}. It was shown that the current-induced insulator-metal transition is due to the formation of in-gap states with only a minor rearrangement of the Mott state \cite{Curcio23}.
In an external electric field, stripe domains with inequivalent octahedral
distortions form \cite{Gauquelin23}. Since the creation of the stripes strongly depends on the direction of the electric field, the interplay between the crystal symmetries and the spin-orbit effect plays a crucial role in this process. 

\begin{figure}[t!]
\centering
\includegraphics[width=9cm,angle=0]{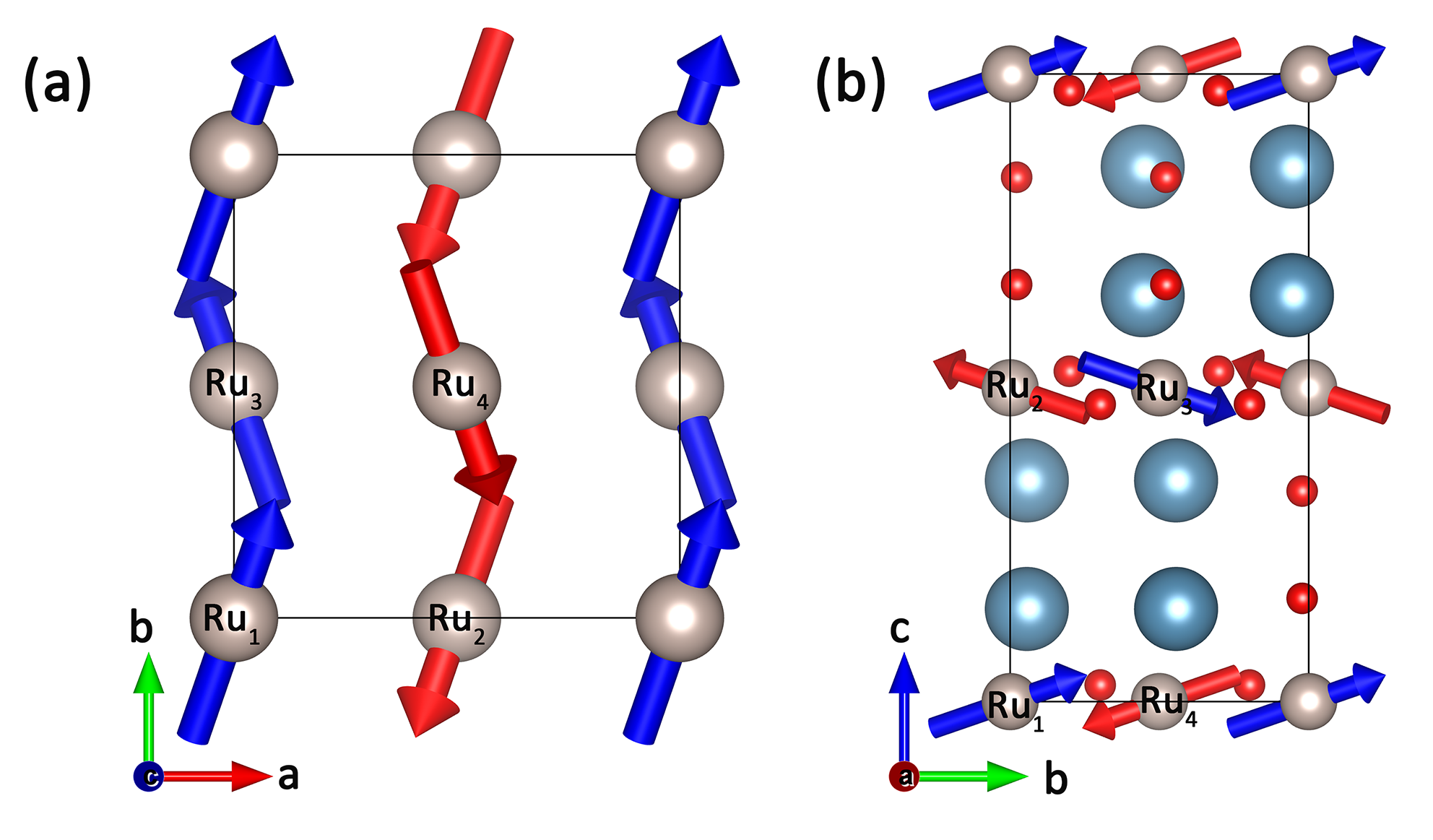}
\caption{Crystal structure of the orthorhombic Ca$_2$RuO$_4$ shown from the (a) top view and (b) side view. Grey, red and blue spheres indicate the Ru, O and Ca atoms, respectively. The arrows represent the spins in the experimental ground state (A-centered) with the N\'eel vector aligned along the b direction. Red and blue arrows represent the spin-up and spin-down in the non-relativistic limit.
For clarity, Ca and O atoms are omitted in panel (a). Spin canting is illustrated with the components along the $a$ and $c$ directions scaled to 35\% of the principal $b$-axis component, to visualize the spin canting better. $a$-, $b$- and $c$-axis, are equivalent to $x$-, $y$- and $z$-axis, respectively.}\label{structure}
\end{figure} 
In this paper, we will demonstrate that Ca$_2$RuO$_4$ hosts spin cantings due to the staggered 
Dzyaloshinskii–Moriya interaction (DMI)\cite{autieri2024staggereddzyaloshinskiimoriyainducingweak} and a relativistic spin-momentum locking for all three spin components. 
Breaking the inversion symmetry with ferroelectric and antiferroelectric distortions generates Rashba- and Weyl-type spin-orbit coupling (SOC). We will unveil the conditions under which the spin-momentum locking and nodal planes can persist. Finally, we show the crystal structure that simulates the stripes hosts the coexistence of two spin-momentum lockings with different d-wave orders. 
The paper is divided as follows: the second section is devoted to the study of the centrosymmetric case in the non-relativistic and relativistic limit, while the third section explores the interplay between altermagnetism and relativistic effects for ferroelectric and antiferroelectric distortions. Section IV describes the properties of the system in the presence of stripes, which produce the combination of two distinct spin-momentum lockings. Lastly, in Section V, the authors draw their conclusions.

\section{The centrosymmetric case}

In this Section, we report the results of our investigation on the altermagnetic properties of the centrosymmetric (CS) Ca$_2$RuO$_4$. The computational framework is the same one used in the literature\cite{ma15196657, Cuono23orbital,Cuono25} and more details are given in the Supplementary Materials. This section is divided into four subsections. In the first subsection, we report the altermagnetic properties of Ca$_2$RuO$_4$ without SOC. In the second subsection, the origin of altermagnetism is explained by using a tight-binding model. In the third and fourth subsections, we include the SOC, reporting on the spin configuration and on the relativistic spin-momentum locking. 

The unit cell of Ca$_2$RuO$_4$ comprises four formula units, with corner-shared RuO$_6$ octahedra arranged in planes, interspersed with alternating CaO layers. The system presents two possible altermagnetic configurations, the A-centered and the B-centered.
The RuO$_6$ octahedra are alternated with clockwise and counterclockwise rotations. The octahedra with the same rotation in the two layers are ferromagnetically coupled in the A-centered mode and antiferromagnetically in the B-centered mode\cite{PhysRevB.72.094104}.
We primarily focus on the A-centered configuration, which is the ground state. In the Supplementary Materials, the features of the altermagnetic band structure in the B-centered mode are reported. The A-centered magnetic configuration is reported in Fig. \ref{structure}(a,b).

\subsection{Non-relativistic Spin-momentum locking}

\begin{figure}[t!]
\centering
\includegraphics[width=7.99cm,angle=0]{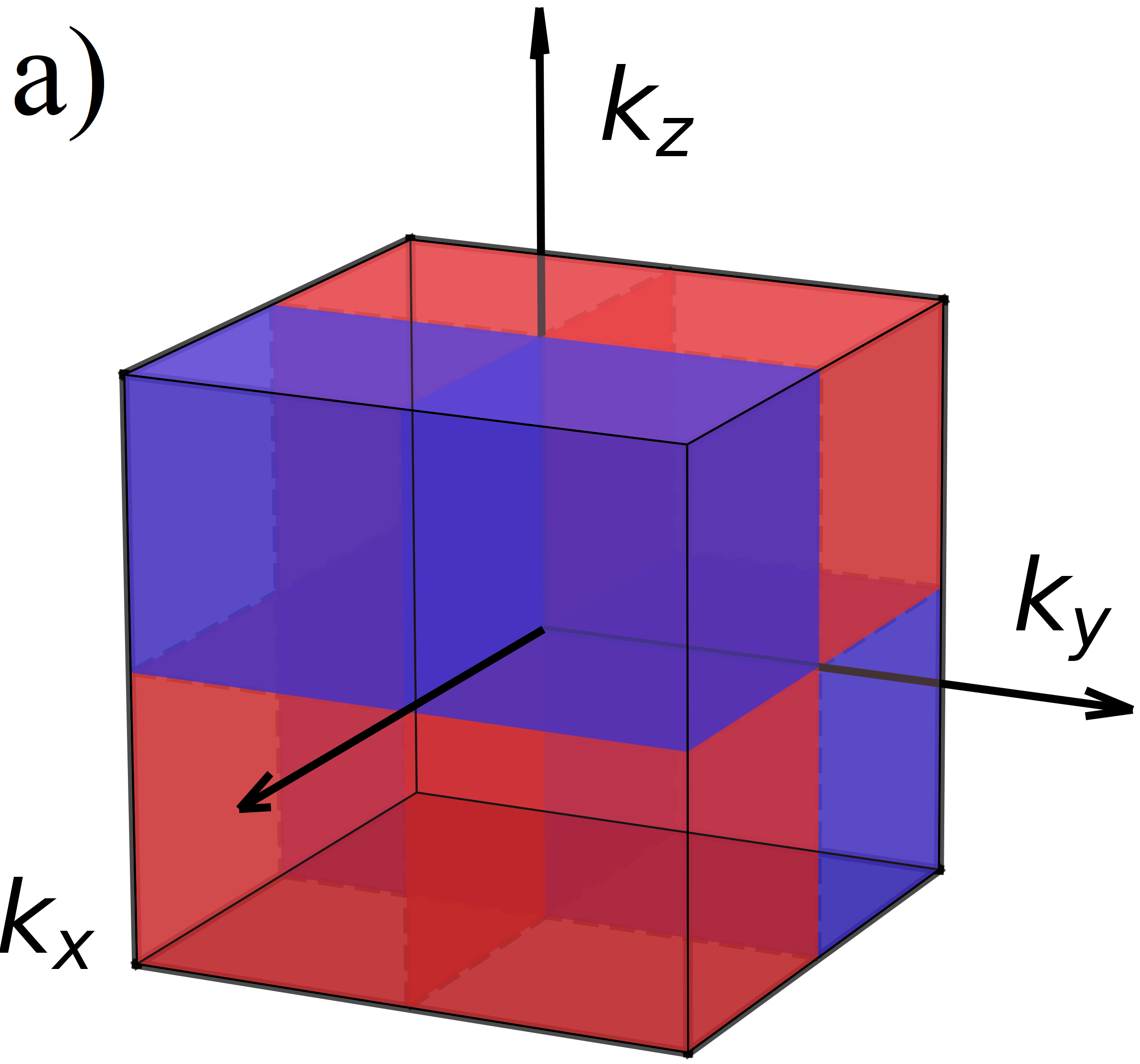}
\includegraphics[width=7.99cm,angle=0]{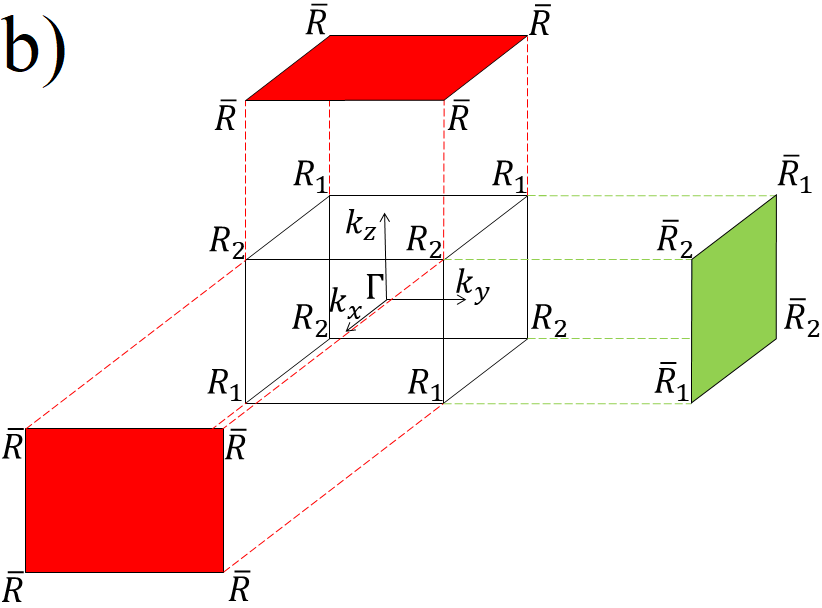}
\caption{a) Spin-momentum locking in the non-relativistic case. Red and blue sectors represent regions of the Brillouin zone with opposite non-relativistic spin-splitting. b) Brillouin zone of  Ca$_2$RuO$_4$ (space group Pbca n. 61) for the A-centered magnetism. In our notation, the high-symmetry points with subscripts 1 and 2 show altermagnetism along the path towards the $\Gamma$ point. We project the bulk Brillouin zone on the principal surfaces (100), (010) and (001). The projected high-symmetry points have an overline. Given the geometrical position of the k-points with opposite non-relativistic spin-splitting, the altermagnetic surface states survived on the (010) surface (colored in green), while the
other two surfaces are blind to AM (colored in red).}\label{BZ}
\end{figure}

The non-relativistic spin splitting does not appear along certain high-symmetry planes of the Brillouin zone (BZ), known as nodal planes. Fig. \ref{BZ}(a) shows the symmetries of the non-relativistic spin-momentum locking in the BZ for the A-centered Ca$_2$RuO$_4$, where red and blue indicate opposite non-relativistic spin-splitting. The non-relativistic spin-momentum locking is planar d-wave, named also P-2 in the literature\cite{Smejkal22beyond}.  Considering the order in the k-space, this can be referred to as a d$_{xz}$ wave. The nodal planes of the non-relativistic spin splitting have equations k$_x$=0 and k$_z$=0. We further emphasize this with the high-symmetry k-point R in Fig. \ref{BZ}(b) 
The subscripts 1 and 2 denote two points in $k$-space with opposite non-relativistic spin splitting. It was already shown that, since altermagnetic spin-splitting occurs along specific k-paths in the 3D Brillouin zone, altermagnetic surface states will manifest on particular surface orientations \cite{D3NR03681B}.
Regarding the electronic surface states, we can project the high-symmetry points on the surface planes. The high-symmetry points projected on the surfaces are labelled via an overline. Given the geometrical configuration of the non-relativistic spin-momentum locking, the altermagnetic surface states survived on the (010) surface, while the other two surfaces are blind to AM.
Therefore, a slab of Ca$_2$RuO$_4$ with (001) orientation does
not host d$_{xz}$ spin-momentum locking. This is valid for the considered unit cell that contains two RuO$_2$ layers; however, for the single layer of RuO$_4$, the system becomes altermagnetic again but with a different spin-momentum locking, which is now d$_{xy}$\cite{D4NR04053H}. As a consequence, we have an altermagnetic odd-even effect for thin films of Ca$_2$RuO$_4$ in which the slabs with an odd number of layers are altermagnets, while the slabs with an even number of layers exhibit PT symmetry.\\

\begin{figure}[t!]
\centering
\includegraphics[width=5.99cm,angle=270]{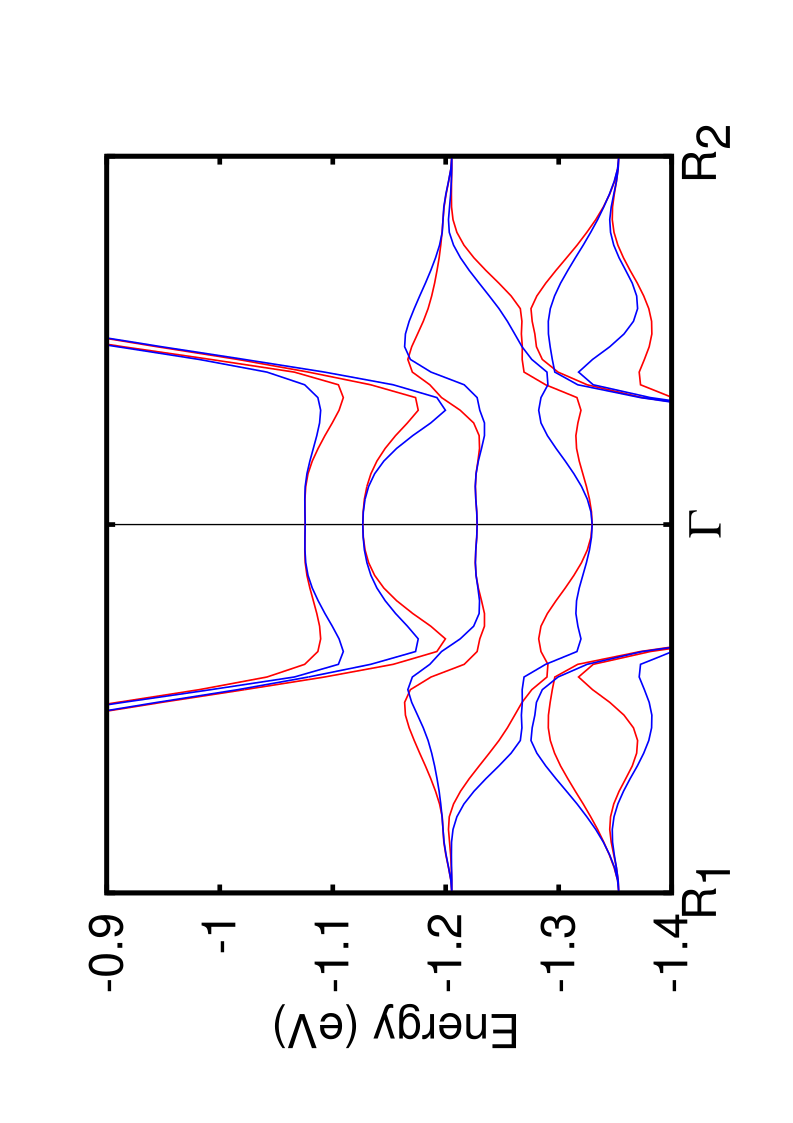}
\caption{Band structure of Ca$_2$RuO$_4$ in the A-centered magnetic phase without SOC along the path R$_1$-$\Gamma$-R$_2$ between -1.4 and -0.9 eV. Blue and red bands represent the spin-up and spin-down channels, respectively. The Fermi level is set to zero energy. The band gap is 0.84 eV. }\label{R1R2_A_centered}
\end{figure}

The size of the non-relativistic spin-splitting can be seen from the band structure along R$_1$-$\Gamma$-R$_2$, as reported in Fig. \ref{R1R2_A_centered}. 
Here, we show a zoom of the band structure in the energy range between -1.4 eV and -0.9 eV; This specific energy window was selected because it hosts the $\gamma_z$ bands, where a pronounced non-relativistic spin-splitting emerges. Such an energy range is experimentally accessible, for example, via resonant inelastic x-ray scattering (RIXS) measurements~\cite{bhartiya2025evidenceelectronicstatesdriving}, in which spectroscopic fingerprints consistent with altermagnetic order can be detected~\cite{Biniskos2025}. It is evident that there is an altermagnetic spin-splitting of the bands along these lines of the BZ. In a previous work, it was shown that in this system, altermagnetism is present in the ${{\gamma}z}$ bands, while the ${xy}$ bands do not present non-relativistic spin-splitting. Thus, the compound shows an exotic type of orbital-selective altermagnetism. \cite{Cuono23orbital}

\begin{table*}[]
\centering
\begin{tabular}{|c|c|c|c|c|c|c|c|c|c|}
\hline
Magnetic Configuration & M$_1$ & M$_2$ & M$_3$ & M$_4$ & M$_{\text{tot}}$ & m$_a$ & m$_b$ & m$_c$ & $\Delta E$ \\
\hline
A-centered with $\Vec{n}$ $\parallel$ $\Vec{a}$ & (m$_a$,m$_b$,m$_c$)   & (-m$_a$,-m$_b$,m$_c$)  & (m$_a$,-m$_b$,m$_c$) & (-m$_a$,m$_b$,m$_c$) & (0,0,4m$_c$) & 1.46 & 0.02 & 0.04 & +0.98 \\
\hline
A-centered with $\Vec{n}$ $\parallel$ $\Vec{b}$ & (m$_a$,m$_b$,m$_c$)  & (-m$_a$,-m$_b$,m$_c$)  & (-m$_a$,m$_b$,-m$_c$)  & (m$_a$,-m$_b$,-m$_c$)  & (0,0,0) & 0.01 & 1.45 & 0.11 & 0 \\
\hline
A-centered with $\Vec{n}$ $\parallel$  $\Vec{c}$ &    (m$_a$,m$_b$,m$_c$) & (m$_a$,m$_b$,-m$_c$)  & (m$_a$,-m$_b$,m$_c$) & (m$_a$,-m$_b$,-m$_c$) & (4m$_a$,0,0) & 0.30 & 0.22 & 1.39 & +36.82  \\
\hline
B-centered with $\Vec{n}$ $\parallel$  $\Vec{a}$ & (m$_a$,m$_b$,m$_c$)   & (m$_a$,m$_b$,-m$_c$)  & (-m$_a$,m$_b$,-m$_c$) & (-m$_a$,m$_b$,m$_c$) & (0,4m$_b$,0) & 1.46 & 0.02 & 0.04 & +2.64   \\
\hline
B-centered with $\Vec{n}$ $\parallel$  $\Vec{b}$ & (m$_a$,m$_b$,m$_c$)  & (m$_a$,m$_b$,-m$_c$)  & (m$_a$,-m$_b$,m$_c$)& (m$_a$,-m$_b$,-m$_c$) & (4m$_a$,0,0)  & 0.01  & 1.45& 0.11  & +0.78  \\
\hline
B-centered with $\Vec{n}$ $\parallel$ $\Vec{c}$    & (m$_a$,m$_b$,m$_c$)  &  (-m$_a$,-m$_b$,m$_c$)& (-m$_a$,m$_b$,-m$_c$) & (m$_a$,-m$_b$,-m$_c$) & (0,0,0) & 0.32 & 0.23 & 1.38 & +37.06  \\
\hline
\end{tabular}
\caption{Magnetic moments and their symmetries, total magnetization, energy difference compared to the ground state and band gap at the $\Gamma$ point of the different magnetic configurations studied for the A- and B-centered phase along different directions of the N\'eel vector ($\Vec{n}$). m$_a$, m$_b$ and m$_c$ are reported in $\mu_B$. $\Delta E$ is reported in meV.}
\label{tab:table1}
\end{table*}

\subsection{Origin of the altermagnetism from the tight-binding model}

We built a minimal tight-binding model for the altermagnetism in the A-centered magnetic phase of Ca$_2$RuO$_4$, to unveil which terms produce the non-relativistic spin-splitting and to understand the origin of the orbital-selective nature of the altermagnetism. The tight-binding model for the t$_{2g}$ orbitals, with both spin majority and spin minority bands considered, is reported in detail in the Supplementary Materials. This tight-binding model is highly complex, comprising several dozen parameters that are all essential for describing the altermagnetic behavior.
In Ca$_2$RuO$_4$, we have a d$_{xz}$ bulk altermagnetism that comes from the interlyer non-diagonal hybridization like $t^{111}_{xz,yz}\sin{k_x}\cos{k_y}\sin{k_z}$. Moreover, Ca$_2$RuO$_4$ hosts a d$_{xy}$ altermagnetic order coming from the intralayer non-diagonal hybridization $t^{110}_{xy,xz}\sin{k_x}\sin{k_y}$, which is hidden in the bulk\cite{hiddenaltermagnetism}, but present in the slab case\cite{D4NR04053H}.
We can observe that the hybridizations of the bulk altermagnetism involve at first order only ${{\gamma}z}$ bands; however, there is no symmetry protecting the proximity of the altermagnetism from the ${{\gamma}z}$ band to the $xy$ bands. Therefore, the orbital selectivity represents a strong suppression of the non-relativistic spin-splitting in the ${xy}$ bands, but it is an accidental characteristic of the bulk A-centered phase. We have observed that the orbital-selective altermagnetism disappears either upon the breaking of structural distortions or in the presence of B-centered magnetism.

\subsection{Spin moments with spin-orbit coupling}

\begin{figure*}[t!]
\centering
\includegraphics[width=16.91cm,angle=0]{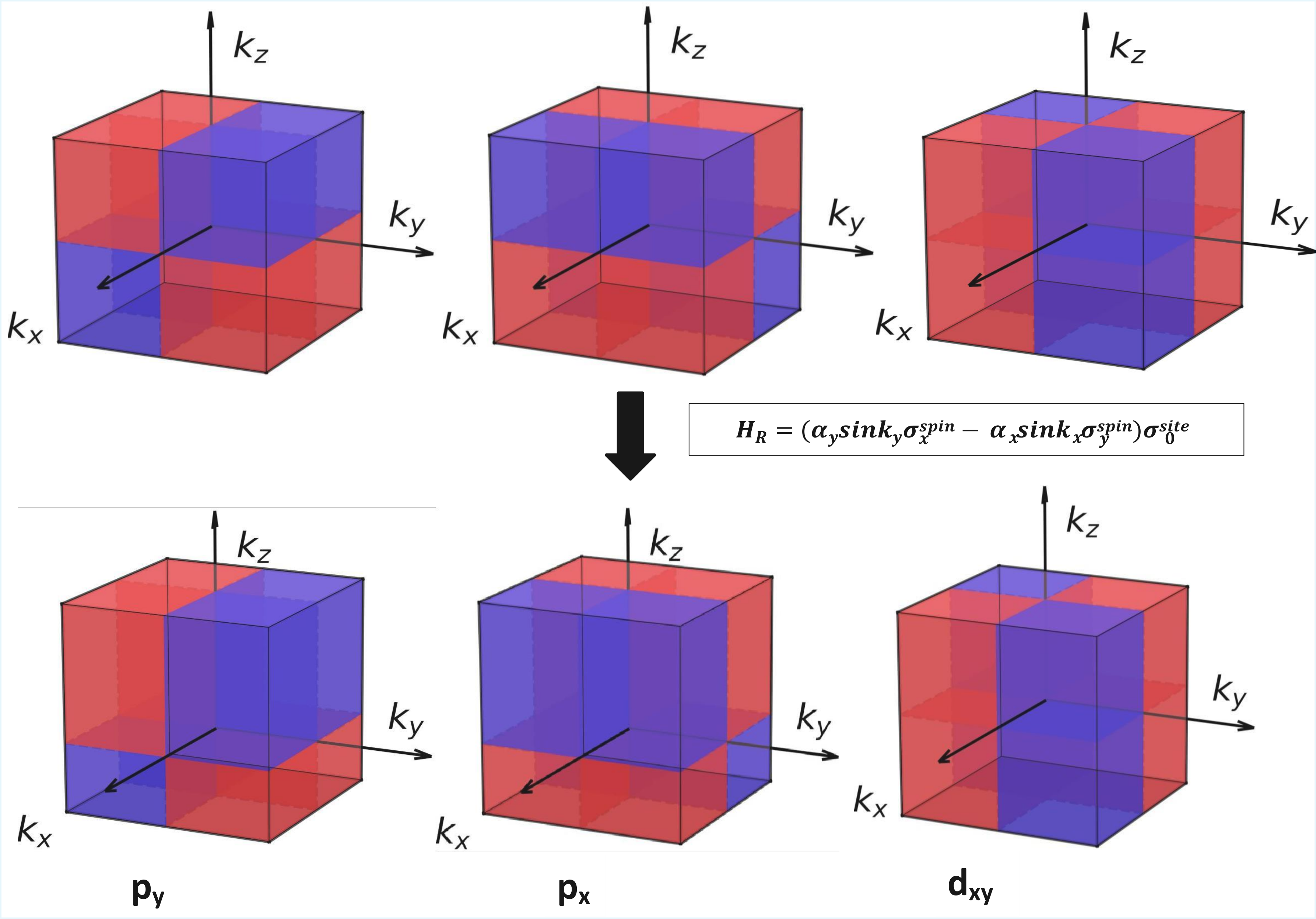} 
\caption{In the top part, relativistic spin-momentum locking of Ca$_2$RuO$_4$ with N\'eel vector along the $y$-axis composed of d$_{yz}$, d$_{xz}$, and d$_{xy}$ for the S$_x$, S$_y$, and S$_z$ components, respectively. Red and blue sectors represent regions of the Brillouin zone with opposite non-relativistic spin-splitting. The S$_y$ component is the dominant component and inherits the non-relativistic spin-momentum locking. 
Once we introduce the Rashba Hamiltonian H$_R$ with an electric field along the z-direction, one symmetry-protected nodal plane is destroyed for the S$_x$ and S$_y$ components, but the spin-momentum locking of the component parallel to the electric field survives. The relativistic spin-momentum locking is composed of p$_y$ for S$_x$, p$_x$ for S$_y$, while the d$_{xy}$ for S$_z$ remains intact.
}\label{RSML} 
\end{figure*}

In this subsection, we will study the spin cantings from the antisymmetric exchanges in the centrosymmetric cases. Upon applying SOC, the relativistic weak ferromagnetism depends on the magnetic order (A-centered or B-centered) and on the N\'eel vector orientation. The results of the total energies correctly predicted the experimental ground state to be the A-centered magnetic order with the N\'eel vector along the $b$-axis. At the same time, there is a tiny difference between the other magnetic phases when the N\'eel vector is in the plane. The phase with N\'eel vector along the $c$-axis is much higher in energy and will not be considered further. The magnetocrystalline anisotropy, defined as the difference between the in-plane and out-of-plane spins per formula unit, is 9.4 meV, which is extremely large. This is, however, compatible with the strong anisotropy of the quasi-2D system and the strong SOC of Ru atoms. 
Our DFT calculations yield magnetic moments of m$_b$ = 1.45 $\mu_B$, m$_a$ = 0.01 $\mu_B$, and m$_c$ = 0.11 $\mu_B$, which closely align with experimental observations, where m$_c\approx$~ 0.1m$_b$ and m$_a\approx$~0\cite{Porter18}. 
Given the positions of the Ru atoms as Ru$_1$ = (0,0,0), Ru$_2$ = (0.5,0,0.5), Ru$_3$ = (0,0.5,0.5), and Ru$_4$ = (0.5,0.5,0) as shown in Fig. \ref{structure}, their magnetic moments in the A-centered configuration with N\'eel vector along the $b$-axis are reported in Table \ref{tab:table1}. As evident from the results, no weak ferromagnetism occurs when the N\'eel vector is along the $b$-axis.
In contrast, when the N\'eel vector is oriented along the $a$- or $c$-axis, weak ferromagnetism emerges along the $c$- and the $a$-axis, respectively.
Numerical data of the magnetic moments and magnetizations are provided in Table \ref{tab:table1} for different N\'eel vector orientations.
The B-centered phase also exhibits spin canting for all N\'eel vector orientations; however, the magnitude and direction of the canting differ between configurations. The two most stable B-centered configurations occur when the N\'eel vector lies in the $ab$ plane, and both exhibit weak ferromagnetism. Among the 6 magnetic configurations, four of them are weak ferromagnets and two of them are pure altermagnetic with zero net magnetization, including the ground state. 

While the $A$-centered configuration with the N\'eel vector oriented along the $b$ axis constitutes the ground state, two competing magnetic states lie within less than 1~meV in energy: the $A$-centered configuration with the N\'eel vector along the $a$ axis and the $B$-centered configuration with the N\'eel vector along the $b$ axis. The ground state is a pure altermagnetic phase with vanishing net magnetization. In contrast, the latter two configurations exhibit weak ferromagnetism, characterized by finite net magnetization components oriented in-plane and out-of-plane, respectively. Recent studies have shown that slight doping of Ca$_2$RuO$_4$ can induce weak ferromagnetism~\cite{Porter2022,stt1-v24p,PhysRevB.88.024413}. It has been proposed that this behavior originates primarily from a doping-driven transition to the $B$-centered phase. Based on our results summarized in Table~\ref{tab:table1}, the $A$-centered and $B$-centered phases can be experimentally distinguished through the direction of the weak ferromagnetic moment.

Looking at the results shown in  Table \ref{tab:table1}, the canted components of the spin can also be larger than 10\% of the total spin moment. This makes Ca$_2$RuO$_4$ one of the altermagnets with the largest deviation from collinearity. 
Despite the extremely tiny non-relativistic spin-splitting, the spin-canting is very large because it depends on the large SOC and not on the size of the non-relativistic spin-splitting. In addition, in oxides hosting altermagnetism, there is a tendency to host a staggered DMI interaction\cite{PhysRevB.86.094413,autieri2024staggereddzyaloshinskiimoriyainducingweak}, which is a first-order spin-orbit driven magnetic interaction rather than higher-order spin-orbit driven interactions\cite{839n-rckn}. While the spin-canting in altermagnets always produces weak ferromagnetism in systems with two atoms per unit cell, the same is not true for altermagnets with four magnetic atoms. The ground state of the Ca$_2$RuO$_4$ provides a counterexample of a canted centrosymmetric altermagnet that exhibits no weak ferromagnetism.

\begin{figure*}[t!]
\centering
\includegraphics[width=16.91cm,angle=0]{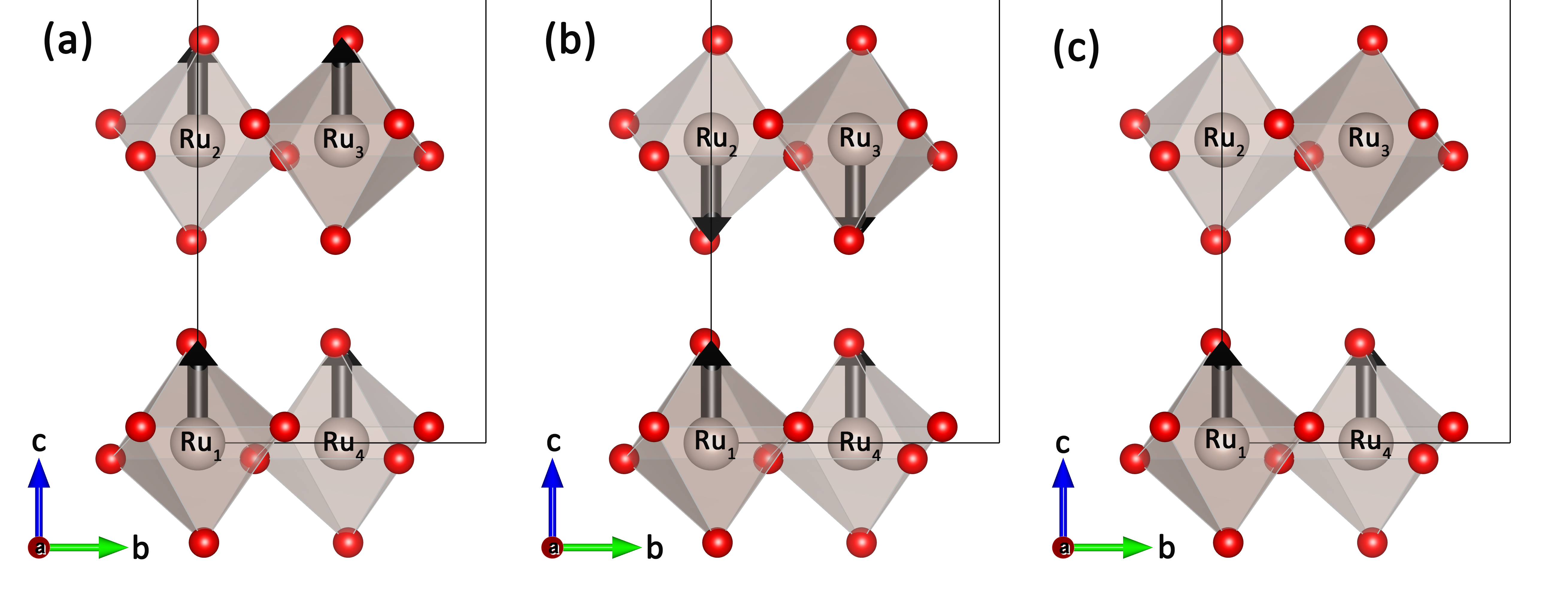} 
\caption{Shift of the Ru-atoms in the (a) ferroelectric-like distortions, (b) antiferroelectric-like distortions, and (c) stripe cases. Red and gray balls represent the oxygen and ruthenium atoms, respectively. The black arrows indicate the displacements of the Ru atoms.  Only the case of shifts along the $z$-axis is shown here. The black lines represent the unit cell of the system.
}\label{Displacements} 
\end{figure*}

\subsection{Relativistic spin-momentum locking for the three spin components}

Moving to the relativistic case, we focus on the A-centered ground state with the N\'eel vector along the $b$-axis (named also $y$-axis). We study the effects of  SOC on the spin-momentum locking.  
The ${xy}$ bands remain Kramers degenerate (not shown); therefore, the SOC preserves the orbital-selective altermagnetism in bulk Ca$_2$RuO$_4$. In the non-relativistic case, the spin-momentum locking is planar d-wave with d$_{xz}$-order. With the N\'eel vector along the $b$-axis, out of the three spin components S$_x$, S$_y$ and S$_z$, the dominant spin component is the S$_y$ component, which will inherit the same spin-momentum locking as in the non-relativistic case. However, in Ca$_2$RuO$_4$ the antisymmetric exchange produced the spin-canting for the S$_x$ and S$_z$ components, and hence both will exhibit spin-momentum locking.
Therefore, Ca$_2$RuO$_4$ exhibits spin–momentum locking in all three spin components, a property that we refer to as relativistic spin–momentum locking\cite{AutieriRSML,hirakida2025multipoleanalysisspincurrents}. As reported in the top part of  Fig. \ref{RSML}, for the S$_x$ component, the spin-momentum locking exhibits d$_{yz}$-wave, while the S$_z$ component exhibits d$_{xy}$-wave. The relativistic spin–momentum locking depends on the direction of the N\'eel vector\cite{AutieriRSML}. In this paper, we consider only the case where the N\'eel vector is aligned along the $b$-axis, as observed experimentally.

Overall, we can write a model Hamiltonian that describes the relativistic spin-momentum locking for Ca$_2$RuO$_4$. Considering a simplified effective single-orbital in terms of Pauli matrices for two spins and two sites, the non-relativistic part with an effective N\'eel vector along the $z$-axis can be described as:
\begin{align} 
\label{eq:H0s}
\mathcal{H}^0  =& \, \, \, \varepsilon(\boldsymbol{k})\sigma_0^{spin}\sigma_0^{site} \\ 
\nonumber 
\mathcal{H}^{\rm AM}_{Sz}  = & {\Delta_z}\sigma_z^{spin}\sigma_z^{site} \\ 
\label{eq:HAMs} & +4t_{am}\sin{k_x}\sin{k_y}\sigma_0^{spin}\sigma_z^{site}  
\end{align}
where t$_{am}$ originates from the octahedral rotations, which are opposite for the sites with opposite spins\cite{autieri2025conditionsorbitalselectivealtermagnetismsr2ruo4}. 
We note that altermagnetism is only present if both the spin-splitting $\Delta_z$ and the hopping parameters t$_{am}$ are not zero. The term 4$t_{am}\sin{k_x}\sin{k_y}$ describes the spin-momentum locking of the S$_z$ component.
For the other two subdominant components, the hopping producing the spin-momentum locking is activated by the SOC $\lambda$ via the antisymmetric exchange. We named $\Delta_x$ and $\Delta_y$ for the relative equations to be 
\begin{equation}\label{eq:Sx}
\mathcal{H}^{\rm AM}_{Sx}=\Delta_x\sin{k_x}\sin{k_z}\sigma_x^{spin}\sigma_z^{site} 
\end{equation}
and 
\begin{equation}\label{eq:Sy}
 \mathcal{H}^{\rm AM}_{Sy}=\Delta_y\sin{k_y}\sin{k_z}\sigma_y^{spin}\sigma_z^{site} 
\end{equation}
Note that the terms in equation (\ref{eq:Sx}) and (\ref{eq:Sy}) do not coincide with the spin-momentum locking of S$_x$ and S$_y$, since the spin-momentum locking of S$_z$ should be also taken into account.\cite{AutieriRSML} The validity of the model Hamiltonian is confirmed by the consistency with DFT results.
The preceding four equations constitute an effective simplified model that reproduces the relativistic spin-momentum locking reported in the top part of Fig.~\ref{RSML}, as demonstrated in the final Section of the Supplementary Materials.  
The use of the effective N\'eel vector along the z-axis in the initial equation (\ref{eq:H0s}) was adopted to make the equation similar to the formalism of the literature paper\cite{Smejkal22,Smejkal22beyond}. Therefore, if we want to capture the experimental properties of Ca$_2$RuO$_4$, we should restrict the analysis to the regime in which the N\'eel vector is aligned along the $y$-axis and the magnetization is stronger along the $y$-axis, corresponding to $\Delta_y \gg \Delta_x, \Delta_z$. The altermagnet Ca$_2$RuO$_4$ hosts three inequivalent spin components exhibiting planar spin-momentum locking.  Its relativistic spin-momentum locking is analogous to that of MnTe$_2$~\cite{Zhu2024}, even if the space groups of the two compounds differ. Additionally, MnTe$_2$ hosts spin-splitting derived from non-collinear magnetism, not qualifying it as a conventional altermagnet.

\section{Breaking the inversion symmetry}

In this Section, we want to study the evolution of the altermagnetic properties of Ca$_2$RuO$_4$ under the breaking of inversion symmetry. We calculated the electronic properties as a function of a shift of the Ru atoms along the $x$-, $y$-, or $z$-axis. Different types of distortions can be considered. A uniform displacement of the Ru atoms across all layers yields a ferroelectric distortion, while alternating displacements produce an antiferroelectric one. 
These two types of structural distortions are shown in Fig. \ref{Displacements}(a,b). 
In the first subsection, we shift the Ru atoms uniformly along the x-, y- and z-directions to create a ferroelectric distortion; this breaks the inversion symmetry and the space group becomes Pca2$_1$ (space group no. 29). In the second subsection, we shift the Ru atoms of different layers in opposite directions along the x-, y-, and z-directions. In this case, the space group is still Pca2$_1$ (space group no. 29) for a shift along y- and z-directions, while it is P2$_1$2$_1$2$_1$ (space group no. 19) for the shift along x. In Fig. \ref{Displacements} (c), we show the stripe case in which we shift the Ru atoms only along one layer and not the other. 

\begin{table}[h!]
    \centering
    \begin{tabular}{|c|c|c|c|c|c|c|c|c|c|}
        \hline
        \multirow{3}{*}{\textbf{Phase}} & \multirow{3}{*}{\textbf{\quad$\Vec{n}$\quad}} & \multirow{1}{*}{\textbf{UD}} &  \multicolumn{3}{c|}{\textbf{AFE}} & \multicolumn{3}{c|}{\textbf{FE}} \\
        \cline{3-9}
        & & & \textbf{x} & \textbf{y} & \textbf{z} & \textbf{x} & \textbf{y} & \textbf{z} \\
        \hline
        \multirow{3}{*}{A-centered} & a & 1.16 & 1.11 & 1.10 & 1.11 & 1.11 & 1.10 & 1.13 \\
        \cline{2-9}
        & b & 1.21 & 1.16 & 1.16 & 1.17 & 1.17 & 1.16 & 1.20 \\
        \cline{2-9}
        & c & 1.20 & 1.14 & 1.11 & 1.17 & 1.14 & 1.11 & 1.18 \\
        \hline
        \multirow{3}{*}{B-centered} & a & 1.19 & 1.14 & 1.13 & 1.15 & 1.14 & 1.13 & 1.18 \\
        \cline{2-9}
        & b & 1.19 & 1.13 & 1.12 & 1.16 & 1.13 & 1.12 & 1.18 \\
        \cline{2-9}
        & c & 1.24 & 1.19 & 1.16 & 1.23 & 1.19 & 1.16 & 1.21 \\
        \hline
    \end{tabular}
    \caption{Energy difference between valence and conduction band extrema at the $\Gamma$ point of the A- and B-phases along different directions of the N\'eel vector with antiferroelectric (AFE) and ferroelectric (FE) displacements along x, y, and z. The reported energies are in eV. }
    \label{tab:band_gap_displacements}
\end{table}

Before presenting the band structure results, we examine the evolution of the energy difference between the valence and conduction band extrema at the $\Gamma$ point, which is proportional to the band gap. The results are reported in Table \ref{tab:band_gap_displacements}. The energy difference between the valence and conduction band extrema at the $\Gamma$ point for the A-centered with N\'eel vector along the $b$-axis is 1.21 eV. This energy difference gets reduced to 1.16-1.17 eV in the case of AFE distortions and further reduces to 1.10-1.11 eV when switching the N\'eel vector from the $b$-axis to the $a$-axis and to 1.11-1.17 eV when switching the N\'eel vector along c. These changes are qualitatively and quantitatively comparable to the case with FE distortions. The deviation in energies for the B-centered is also of the same range as for the A-centered. 
The band gap for the undistorted A-centered phase with the N\'eel vector along the $b$-axis is 0.89 eV. Antiferroelectric distortions reduce the band gap with a larger effect when weak ferromagnetism is present. For instance, the A-centered phase with the N\'eel vector along the $b$-axis with antiferroelectric distortions has a band gap of 0.84 eV for AFE distortions along the $x$-axis, 0.86 eV along the $y$-axis, and 0.87 eV along the $z$-axis. Therefore, all these ferroelectric and antiferroelectric distortions produce a reduction of the band gap. The changes of the band gap after FE distortions are quantitatively similar to AFE and they are reported in Table \ref{tab:band_gap_displacements}.\\

\begin{figure*}[t!]
\centering
\includegraphics[width=6.1cm,angle=0]{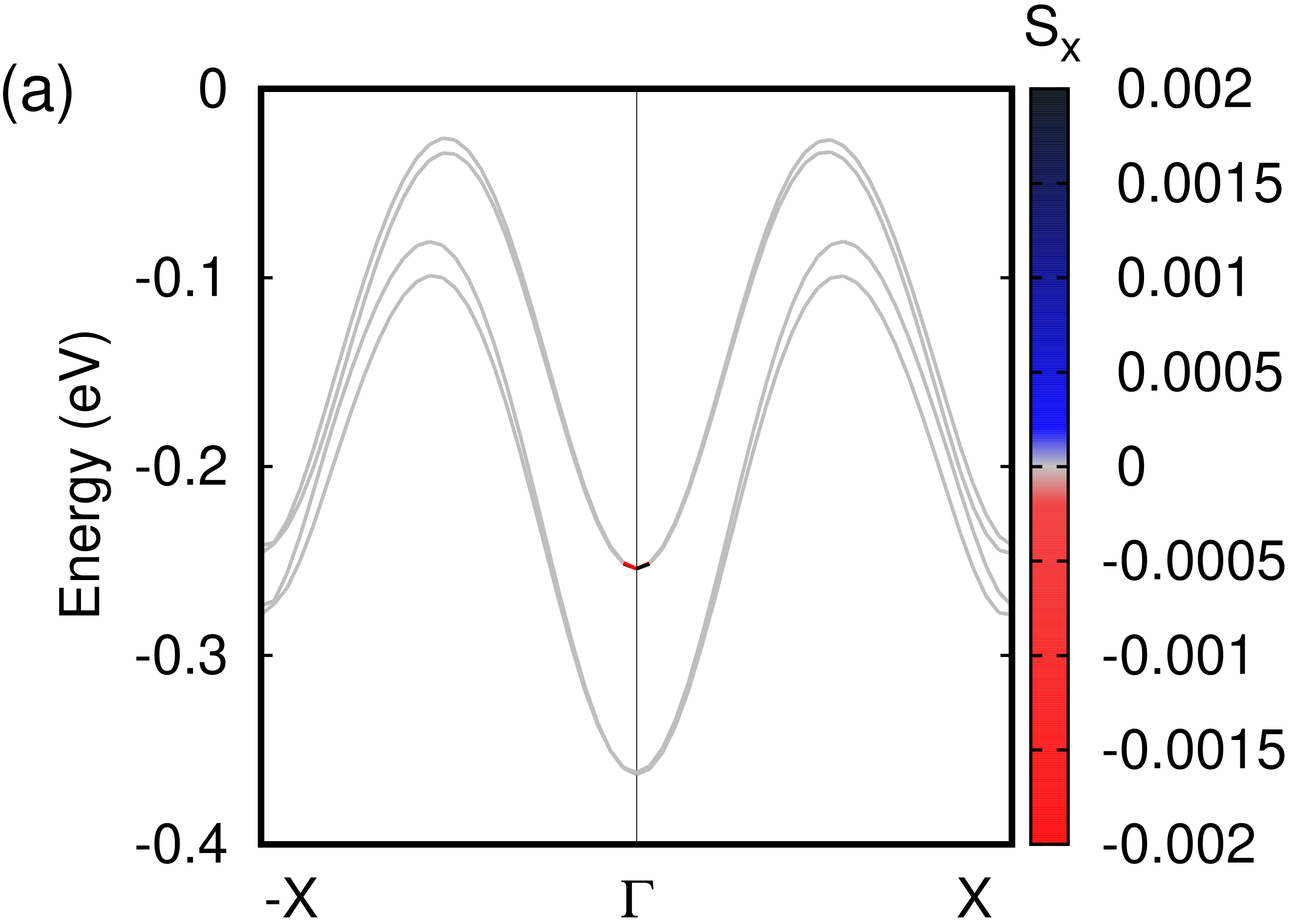}
\includegraphics[width=5.8cm,angle=0]{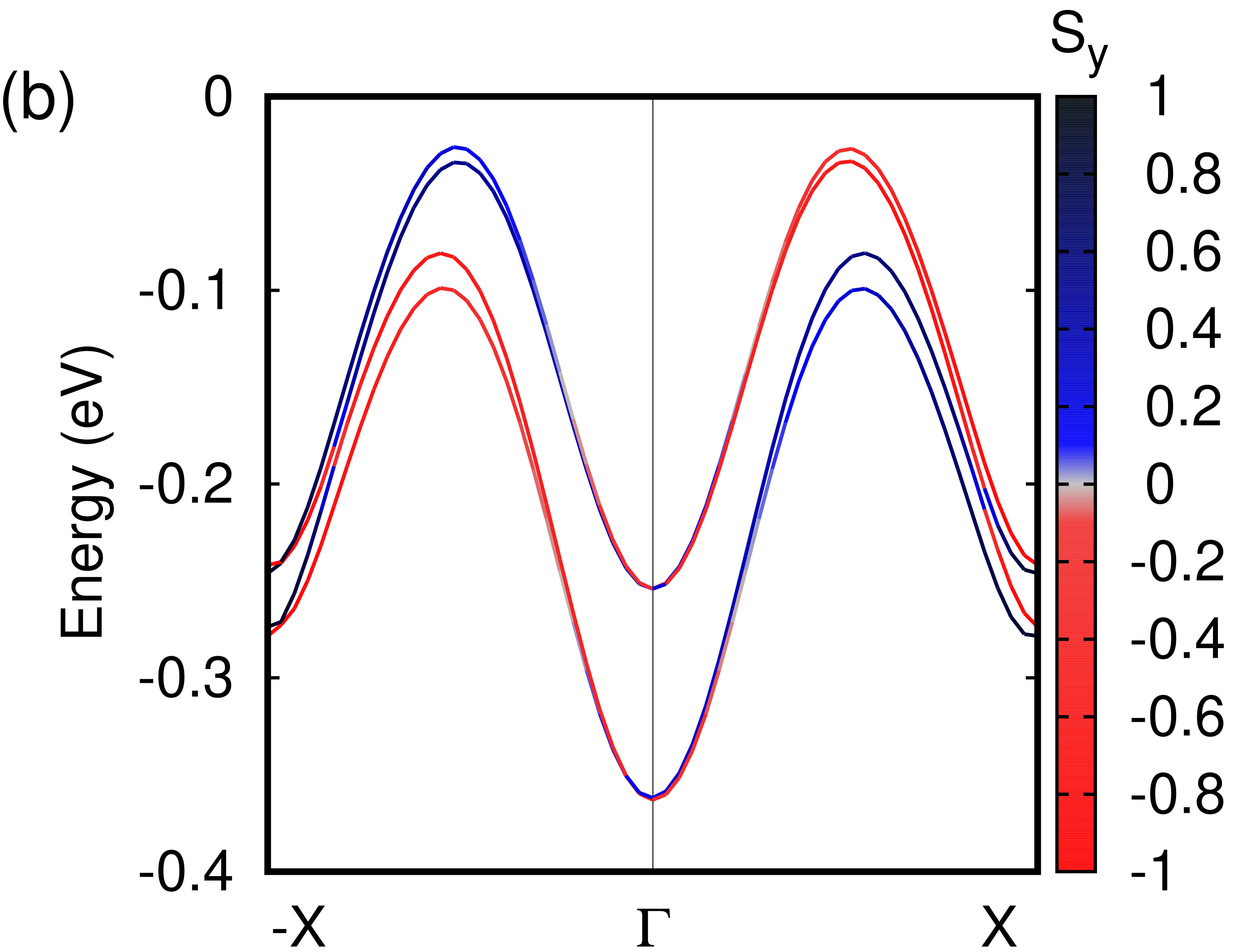}
\includegraphics[width=5.8cm,angle=0]{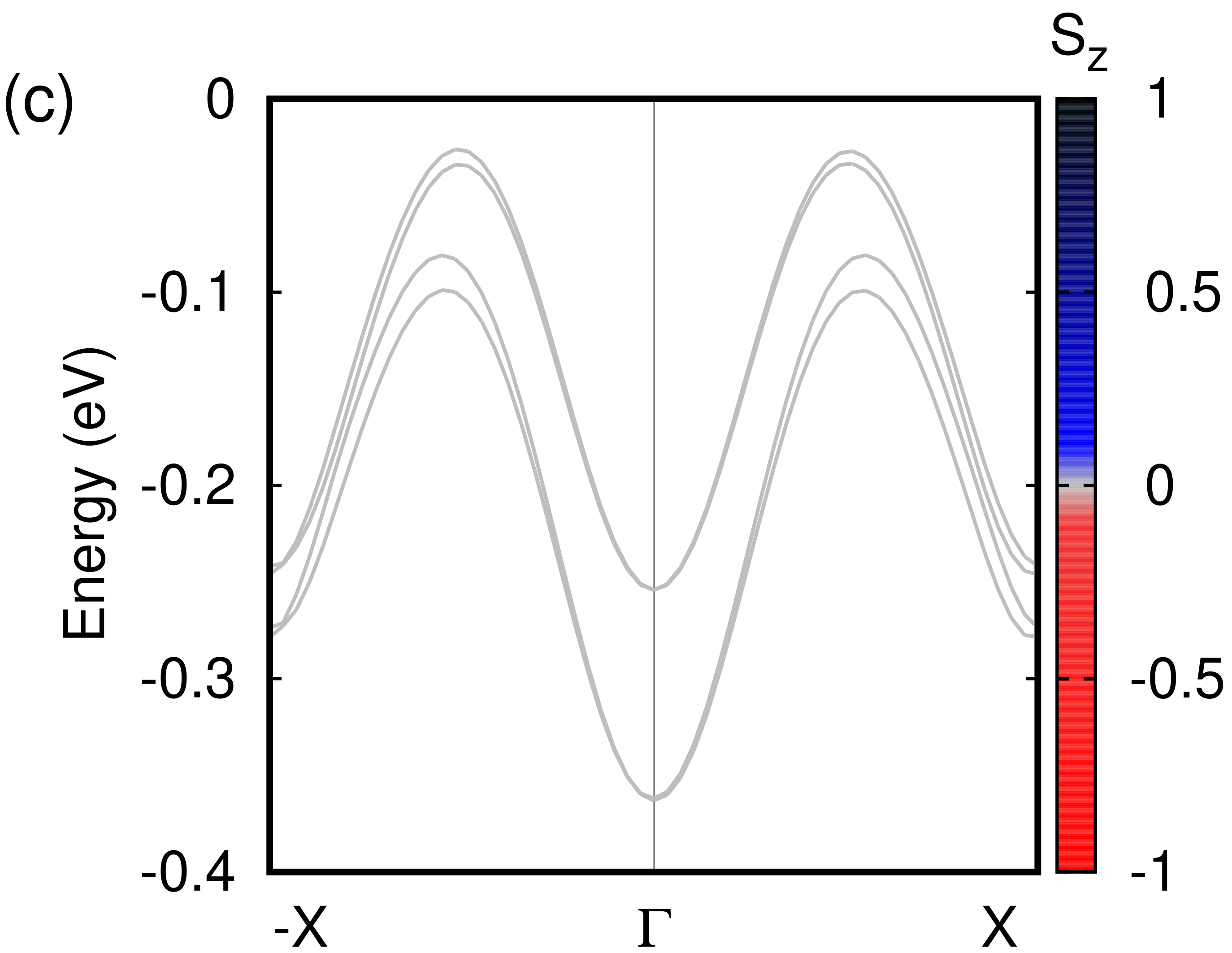}
\caption{Band structures along the $\overline{X}$-$\Gamma$-X direction of the A-centered phase with N\'eel vector along the $b$-axis and ferroelectric distortions along the $z$-axis for the (a) S$_x$ component, (b) S$_y$ component, and 
(c) S$_z$ component.}
\label{figureFEzX} 
\end{figure*}
\begin{figure*}[t!]
\centering
\includegraphics[width=4.5cm,angle=270]{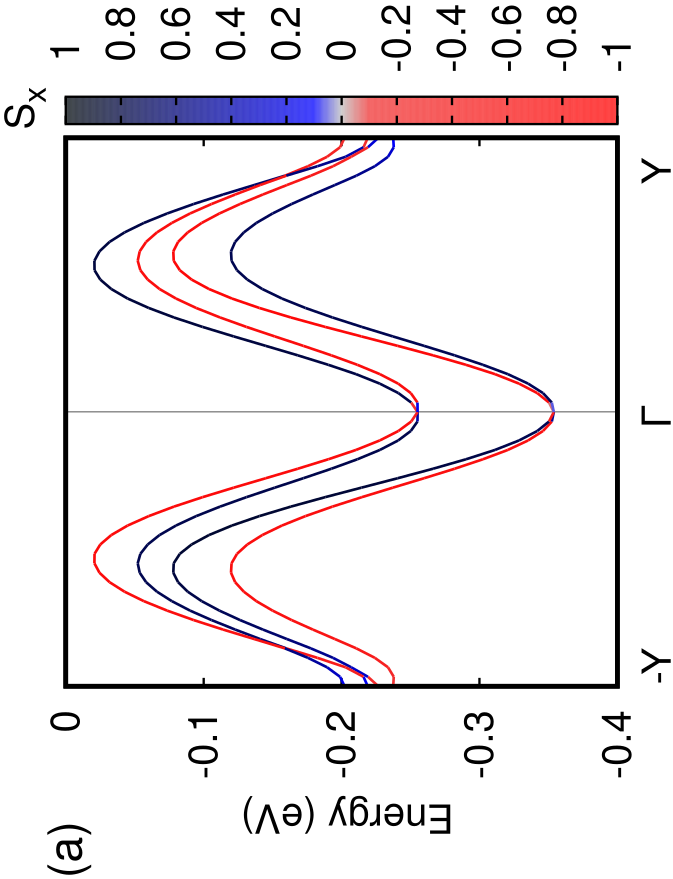}
\includegraphics[width=4.5cm,angle=270]{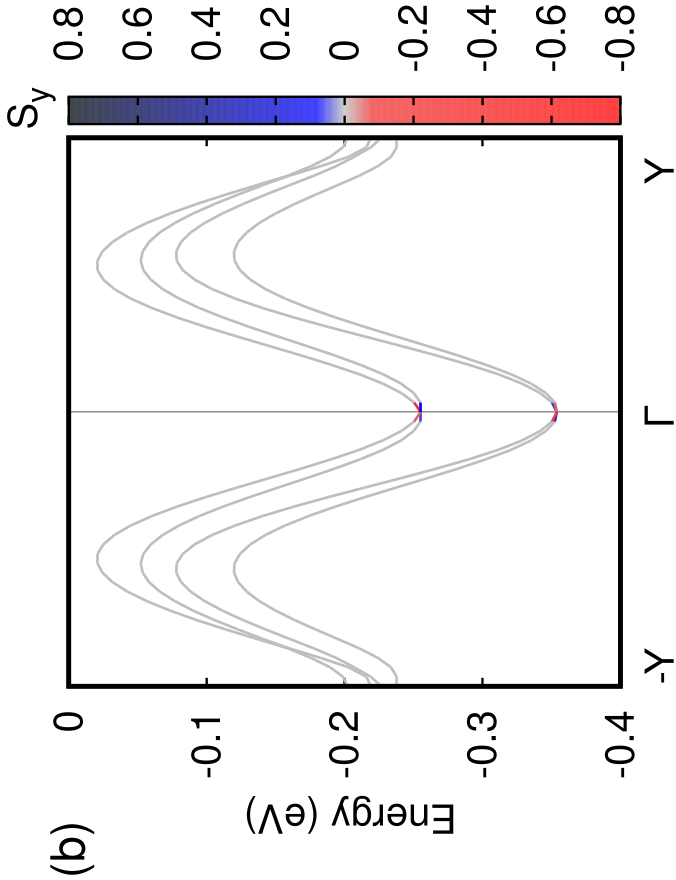}
\includegraphics[width=4.5cm,angle=270]{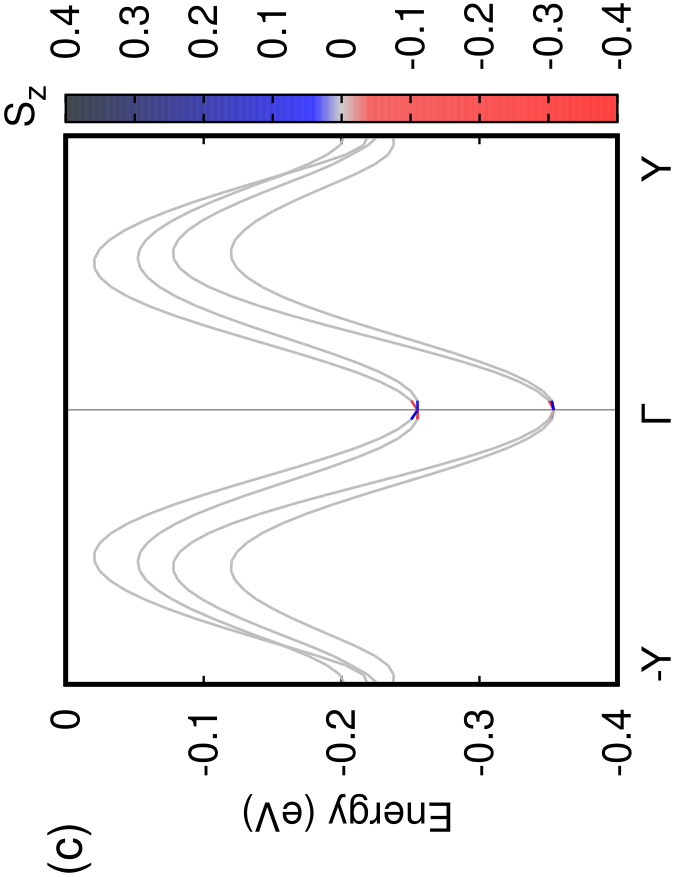}
\caption{Band structures along the $\overline{Y}$-$\Gamma$-Y direction of the A-centered phase with N\'eel vector along the $b$-axis and ferroelectric distortions along the $z$-axis for the (a) S$_x$ component, (b) S$_y$ component, and 
(c) S$_z$ component.
}\label{figureFEzY} 
\end{figure*}

\subsection{Analytic results on the interplay between Rashba and altermagnetism}

We will demonstrate that the relativistic Rashba spin-splitting due to an electric field does not affect the altermagnetic spin-splitting of the spin component parallel to the electric field. Let us consider again the Hamiltonian of an altermagnet with one orbital and two sites given by the sum of the equations (\ref{eq:H0s}-\ref{eq:Sy}). 
The Rashba Hamiltonian for an electric field parallel to the $z$-axis can be written as:
\begin{equation}\label{eq:Rashba}
\mathcal{H}_{R}=({\alpha_y}\sin{k_y}\sigma_x^{spin}-{\alpha_x}\sin{k_x}\sigma_y^{spin})\sigma_0^{site}
\end{equation}
where ${\alpha_y}$ and ${\alpha_x}$ are the coefficients which determine the strength of the Rashba SOC, these coefficients could be quantitatively determined by fitting the relativistic band structure\cite{PhysRevB.109.115141}. 
The eigenvalues of the relativistic altermagnet with the Rashba Hamiltonian (\ref{eq:Rashba}) are given by:
\begin{widetext}
\begin{align} \nonumber
E^{1,\pm}=\varepsilon(\boldsymbol{k})+ 4t_{am} \sin(k_{x}) \sin(k_{y})\\
\pm\sqrt{\,(\alpha_y\sin{k_y}+\Delta_x\sin{k_x}\sin{k_z})^{2} + (\alpha_x\sin{k_x}+\Delta_y\sin{k_y}\sin{k_z})^{2} + \Delta_{z}^{2}}
\end{align}
\begin{align} \nonumber
E^{2,\pm}=\varepsilon(\boldsymbol{k})- 4t_{am} \sin(k_{x}) \sin(k_{y})\\
\pm\sqrt{\,(\alpha_y\sin{k_y}-\Delta_x\sin{k_x}\sin{k_z})^{2} + (\alpha_x\sin{k_x}-\Delta_y\sin{k_y}\sin{k_z})^{2} + \Delta_{z}^{2}}
\end{align}
\end{widetext}
where the eigenvalues $E^{1,\pm}$ are related to  site 1 for the majority (-) and minority spins (+), while $E^{2,\pm}$ are the analogues for site 2.
We can consider a generic k-point P$_1$=($\overline{k_x}$,$\overline{k_y}$,$\overline{k_z}$) and its mirror symmetric counterpart with respect to the nodal plane k$_x$=0 P$_2$=(-$\overline{k_x}$,$\overline{k_y}$,$\overline{k_z}$).
The energies of the minority spins on the two sites are E$^{1-}(P_1)$ and E$^{2-}(P_1)$ and they have opposite S$_z$ component. 
The mirror operation would map E$^{1-}(P_1)$ and E$^{2-}(P_1)$ in E$^{2-}(P_2)$ and E$^{1-}(P_2)$, exchanging the spins and reversing the sign of the spin-splitting for the S$_z$ component. Therefore, the spin–momentum locking changes sign with respect to $k_x$ assuming $\varepsilon(P_1)$=$\varepsilon(P_2)$. A similar analysis can be carried out for $k_y$; as a result, the spin–momentum locking of $S_z$ remains of $d_{xy}$ type even in the presence of the Rashba interaction. Consequently, the Rashba-induced relativistic spin splitting does not affect the spin–momentum locking of the spin component orthogonal to the electric field. Along the nodal planes of the altermagnet, only the Rashba contribution persists, effectively mimicking a relativistic $p$-wave–like spin–momentum locking. 
Without the breaking of inversion symmetry, the nodal plane hosts two degenerate bands with opposite spin components. Once the Rashba interaction rises due to an electric field along the $z$-axis, the bands split and one observes a non-zero S$_x$ component along $\Gamma$ to Y and a non-zero S$_y$ component along $\Gamma$ to X.

\begin{figure*}[t!]
\centering
\includegraphics[width=4.4cm,angle=270]{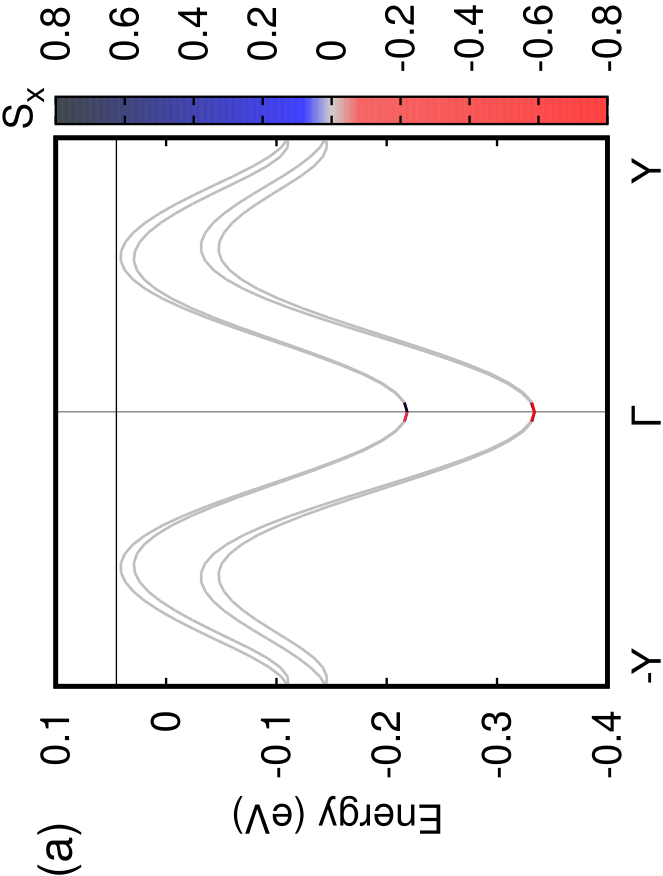}
\includegraphics[width=4.4cm,angle=270]{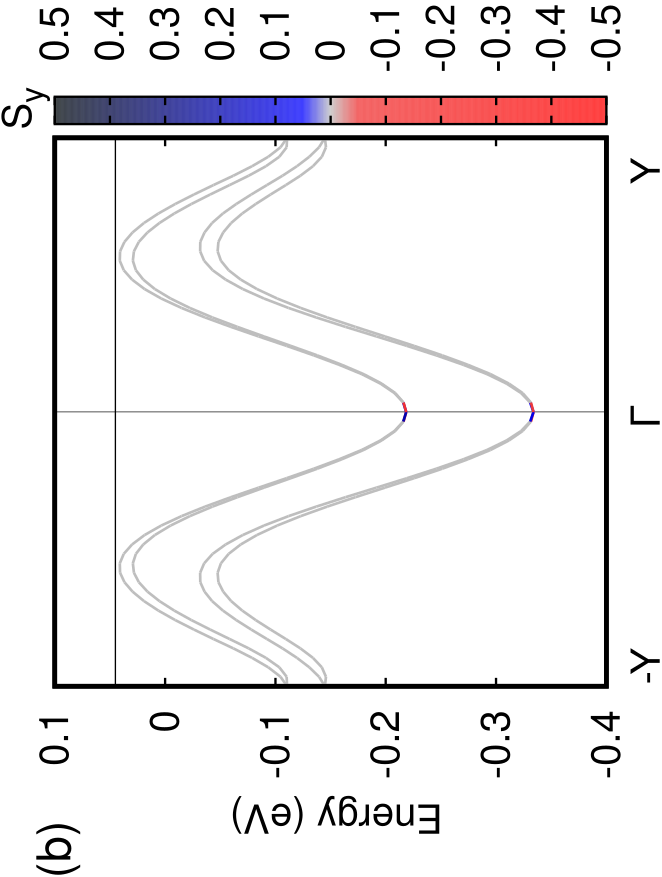}
\includegraphics[width=4.4cm,angle=270]{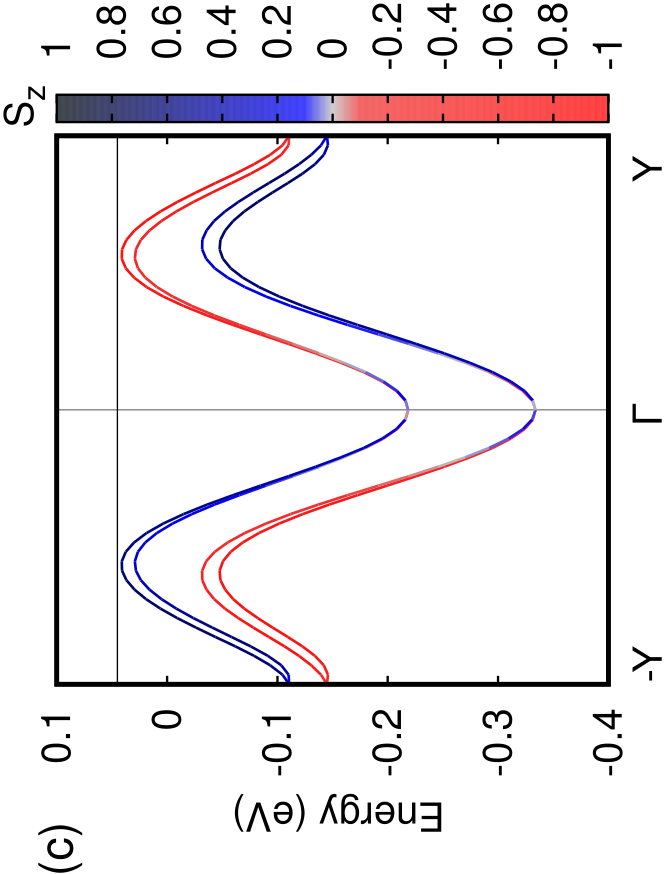}
\caption{Band structures along the $\overline{Y}$-$\Gamma$-Y direction of the A-centered phase with N\'eel vector along the $b$-axis and ferroelectric distortions along the $x$-axis for the (a) S$_x$ component, (b) S$_y$ component, and 
(c) S$_z$ component.
}\label{figureFExY} 
\end{figure*}
\begin{figure*}[t!]
\centering
\includegraphics[width=4.6cm,angle=270]{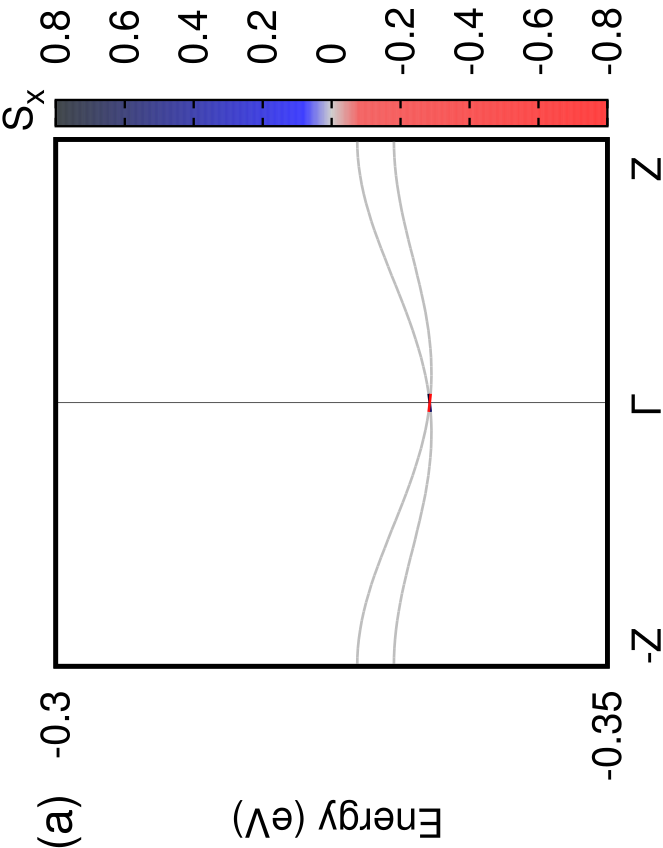}
\includegraphics[width=4.6cm,angle=270]{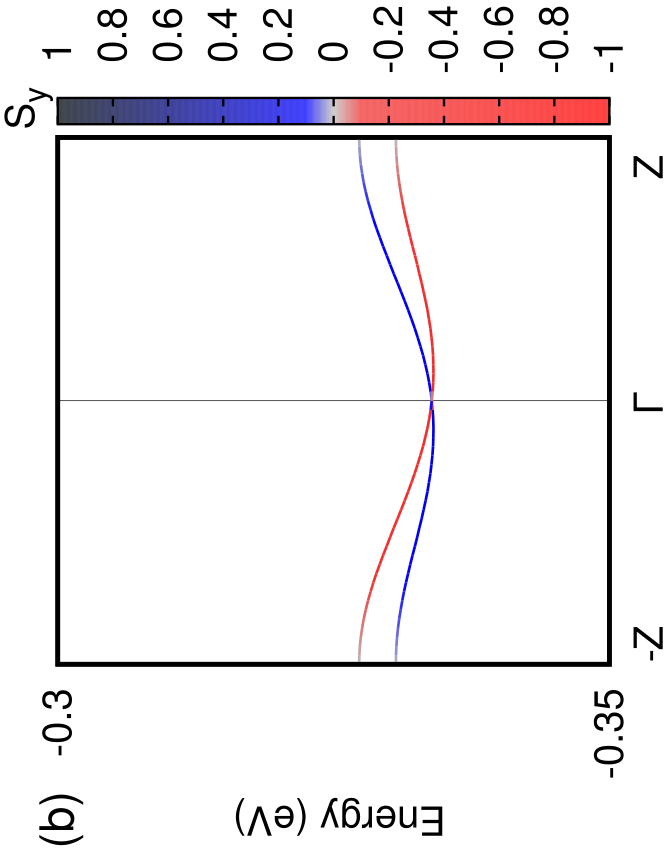}
\includegraphics[width=4.6cm,angle=270]{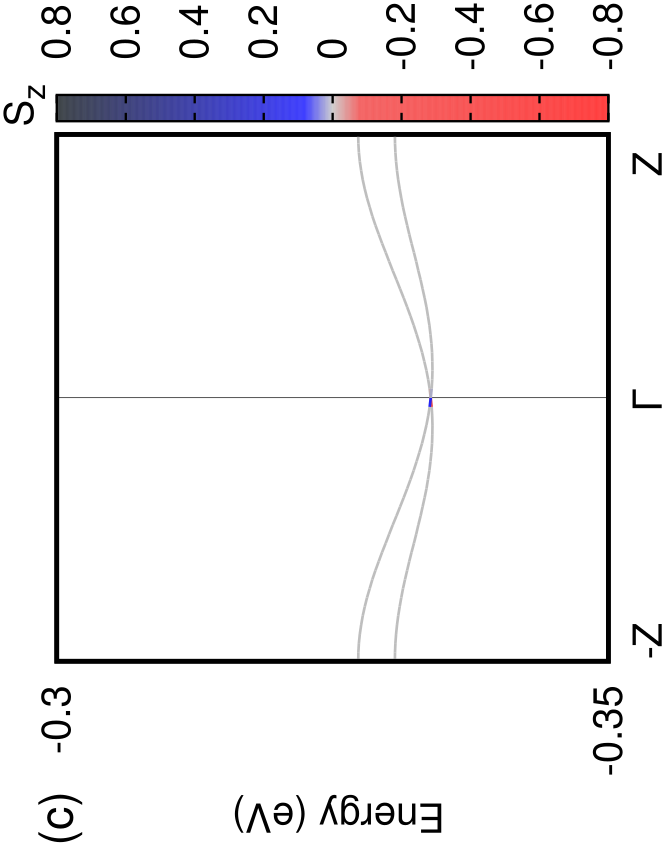}
\caption{Band structures along the $\overline{Z}$-$\Gamma$-Z direction of the A-centered phase with N\'eel vector along the $b$-axis and ferroelectric distortions along the $x$-axis for the (a) S$_x$ component, (b) S$_y$ component, and 
(c) S$_z$ component.
}\label{figureFExZ} 
\end{figure*}

Let us reconsider the relativistic spin-momentum locking at the top of Fig. \ref{RSML}.
For the S$_x$ component, the Rashba part $\alpha_y\sin{k_y}$ favors the sign of the spin component for $k_y>0$ and favors the opposite sign for 
$k_y<0$. For the S$_y$ component, the Rashba part $\alpha_x\sin{k_x}$ favors one sign for $k_x>0$ and the opposite one for $k_x<0$. The results of the electric field along the z-axis are reported in the bottom part of Fig. \ref{RSML}; the nodal plane k$_z$=0 is broken for both the spin-momentum locking of the S$_x$ and S$_y$ components.
A small electric field would produce the results in the bottom part of Fig. \ref{RSML} as a small perturbation with respect to the centrosymmetric case in the top part of Fig. \ref{RSML}. However, a sufficiently large electric field can completely polarize the bands, producing one half of the Brillouin zone that is fully polarized up and the other half that is fully polarized down.
These results were confirmed using the model Hamiltonian.
In principle, we can argue that, despite not being protected by the crystal symmetry, we have a nodal plane at a constant value of k$_z\propto\alpha_y$ for S$_x$ and k$_z\propto\alpha_y$ for S$_y$; however, such values of $\alpha_x$ and $\alpha_y$ depend on the given band and given spin component, therefore we have no nodal plane at constant k$_z$.
Therefore, the spin-momentum locking of $S_x$ and $S_y$ components becomes p$_y$ and p$_x$, respectively, while, as demonstrated before, the spin-momentum locking of S$_z$ persists.
In summary, the relativistic spin-momentum locking is composed of p$_y$, p$_x$ and d$_{xy}$ for S$_x$, S$_y$ and S$_z$ components, respectively.
We should mention that the rise of the p-wave spin-momentum locking is not guaranteed in any Rashba altermagnet, since it depends on the relativistic spin-momentum locking in the centrosymmetric case. As a counterexample, consider a system in which a given spin component is associated with a $d_{xz}$-wave spin-momentum locking in the centrosymmetric case, while the Rashba interaction introduces a $p_y$-wave spin-momentum locking. In this case, there is no plane in $k$-space where both spin splittings vanish simultaneously, and therefore no relativistic $p$-wave spin-momentum locking emerges.

Despite the breaking of the spin-momentum locking, the Rashba interaction does not induce any weak ferromagnetism; therefore, the Rashba-type spin-orbit breaks the spin-momentum locking for two spin-components but preserves the vanishing magnetization for all three components.
By employing the developed model, we can easily study these relativistic effects by plotting the band structure on the nodal planes. A numerical evaluation of this model is provided in the Supplementary Materials. Moreover, the same analytical framework can be extended beyond the Rashba Hamiltonian to more complex systems, such as the persistent spin helix\cite{tenzin2025persistentspintexturesaltermagnetism} involving $S_x$ and $S_y$, which arises from inversion-symmetry breaking induced by an electric field applied along the $z$-axis.

\subsection{Band structures with spin-orbit coupling and ferroelectric distortions}

\begin{figure*}[t!]
\centering
\includegraphics[width=4.5cm,angle=270]{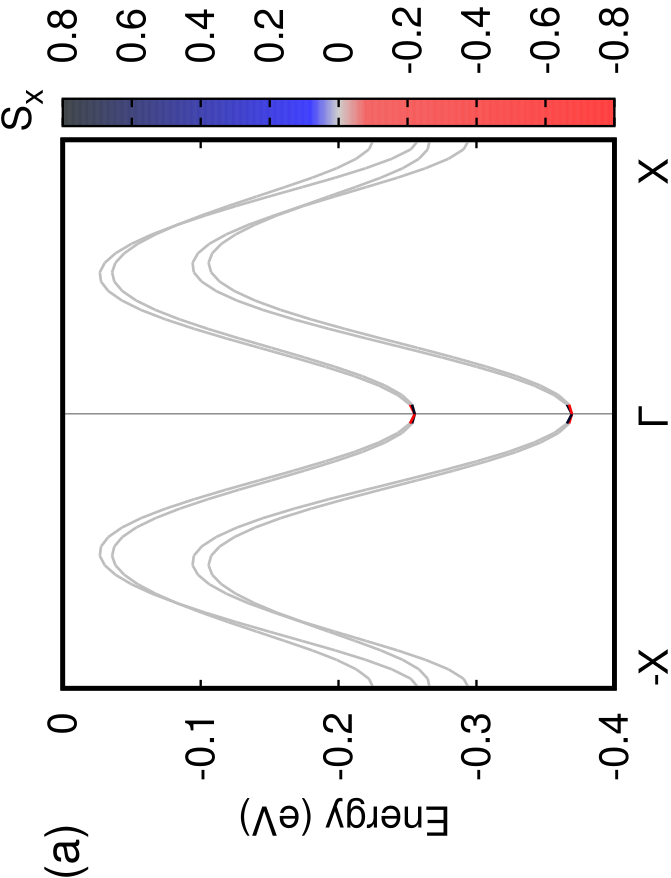}
\includegraphics[width=4.5cm,angle=270]{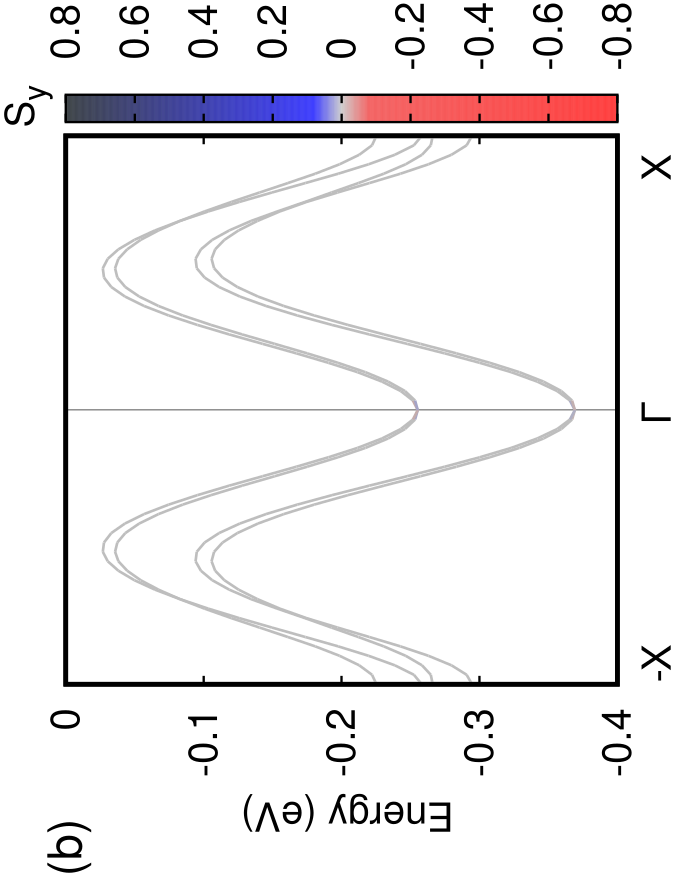}
\includegraphics[width=4.5cm,angle=270]{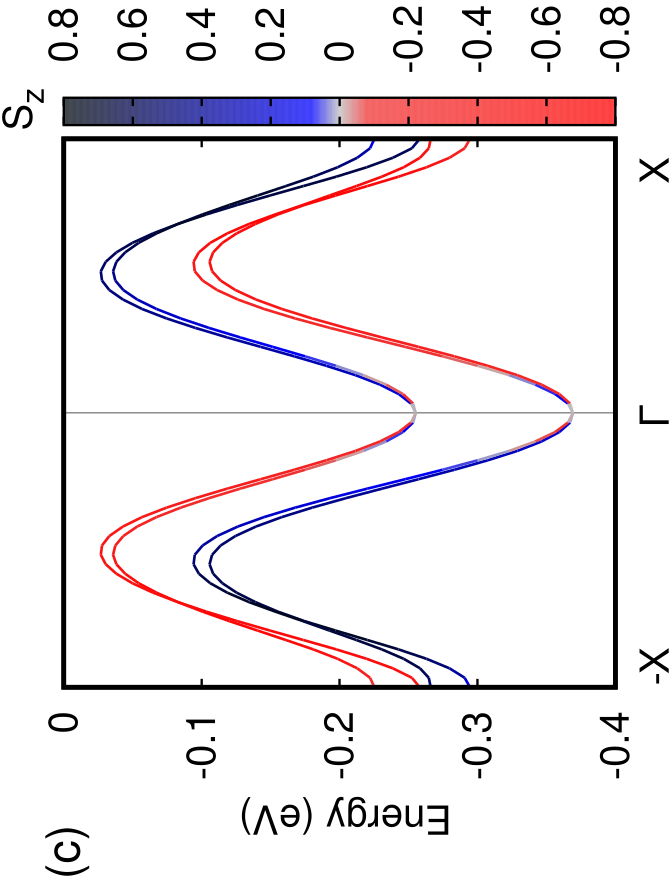}
\caption{Band structures along the $\overline{X}$-$\Gamma$-X direction of the A-centered phase with N\'eel vector along the $b$-axis and ferroelectric distortions along the $y$-axis for the (a) S$_x$ component,  (b) S$_y$ component, and 
(c) S$_z$ component.
}\label{figureFEyX} 
\end{figure*}
\begin{figure*}[t!]
\centering
\includegraphics[width=4.5cm,angle=270]{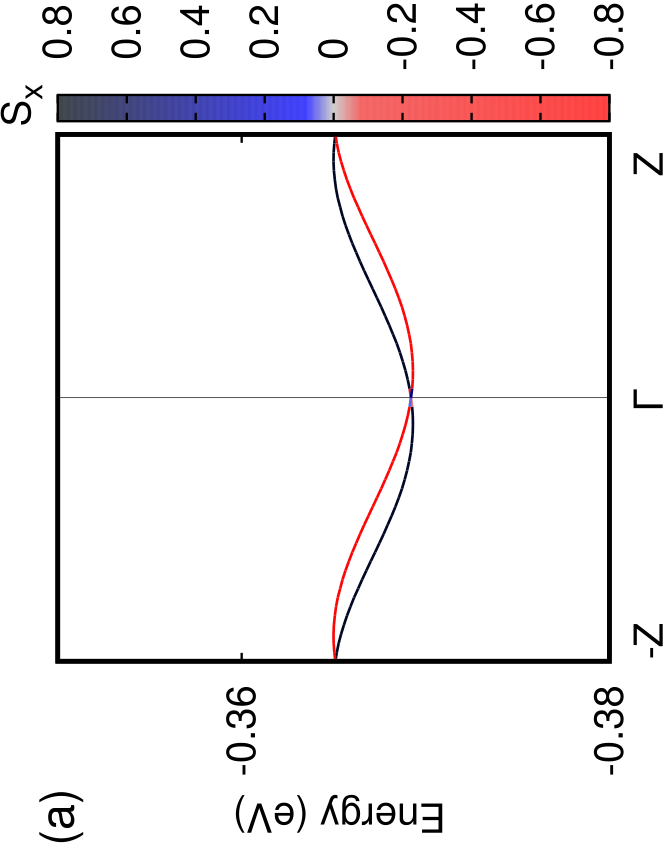}
\includegraphics[width=4.5cm,angle=270]{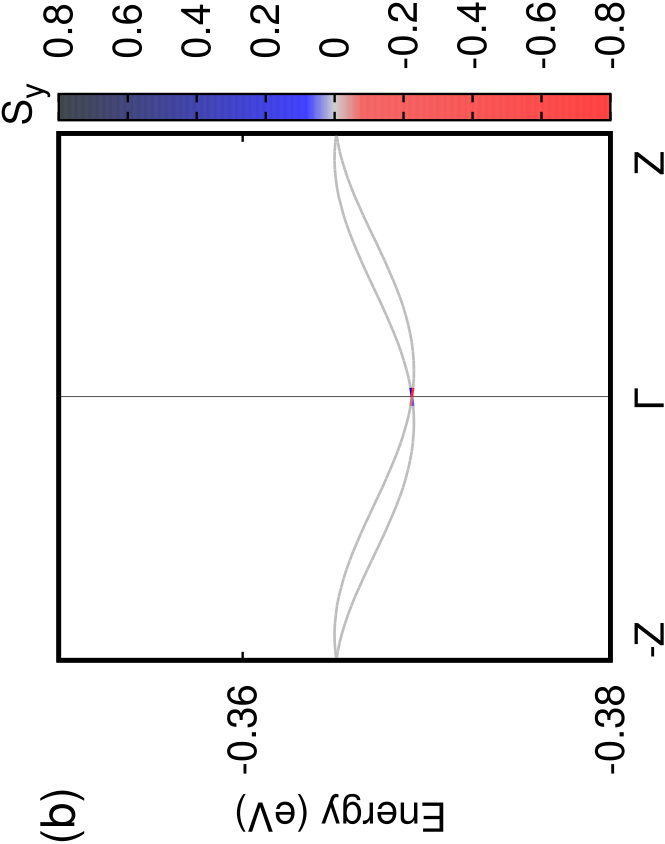}
\includegraphics[width=4.5cm,angle=270]{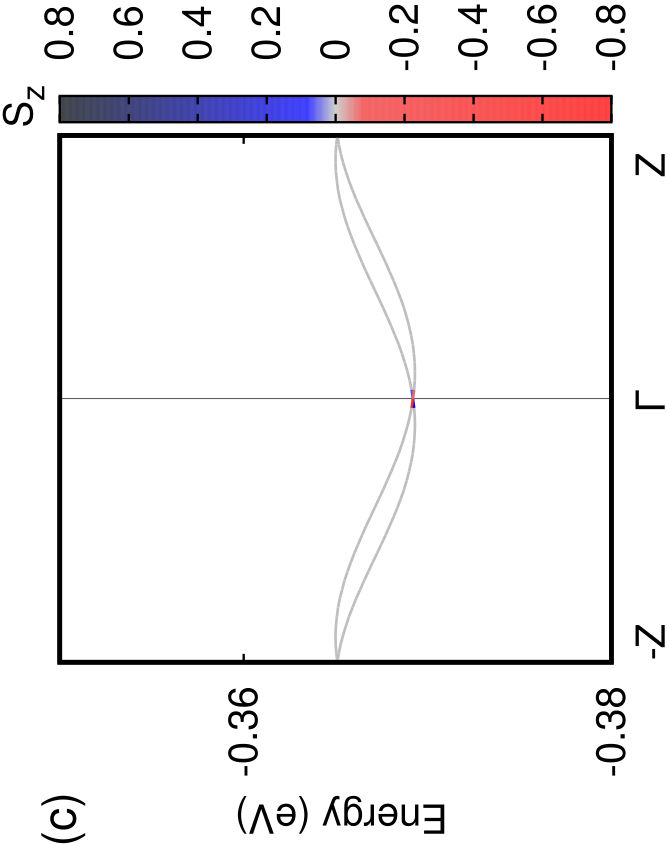}
\caption{Band structures along the $\overline{Z}$-$\Gamma$-Z direction of the A-centered phase with N\'eel vector along the $b$-axis and ferroelectric distortions along the $y$-axis for the (a) S$_x$ component,  (b) S$_y$ component, and 
(c) S$_z$ component.
}\label{figureFEyZ} 
\end{figure*}

Moving to the DFT results, when we apply the uniform displacement of the Ru atoms for the ground state with A-centered magnetism and N\'eel vector along the $b$-axis, we do not observe any change in the spin-momentum locking of the non-relativistic phase and we do not observe any major change in the weak ferromagnetism. In both cases of ferroelectric and antiferroelectric distortions, no qualitative changes are observed in the non-relativistic band structure and in the symmetries of the magnetic moments reported in Table \ref{tab:table1}. The major effect in the non-relativistic band structure is the removal of the orbital-selective altermagnetism; therefore, the non-relativistic altermagnetic properties are intact. On the contrary, in the relativistic case, the shift of the Ru atoms breaks inversion symmetry and introduces the Rashba and/or Weyl effect in the system. 

First, we analyze the shift of Ru atoms parallel to the N\'eel vector equivalent to an electric field along the y-axis (E$_y$). The consequent Rashba effect reads as:
\begin{equation}
H_{R}=({\alpha_y}\sin{k_y}\sigma^{spin}_x-{\alpha_x}\sin{k_x}\sigma^{spin}_y)\sigma_0^{site}
\end{equation}
where the component S$_z$ is not involved in the Rashba effect in this case. The effect of the Rashba interaction along the nodal planes and the band structures are reported in Figs.~\ref{figureFEzX} and \ref{figureFEzY}. 
Figs. \ref{figureFEzX}(b) and \ref{figureFEzY}(a) prove the destruction of the nodal plane k$_z$=0 for the spin-momentum locking for S$_y$ and S$_x$, respectively. At k$_z$=0, there is no longer a nodal plane for the spin–momentum locking of the S$_x$ and S$_y$ components. However, there is still another nodal plane intact; therefore, the S$_x$ and S$_y$ components will have p$_y$ and p$_x$ components, respectively.
The absence of spin spectral weight of S$_x$ at k$_y$=0 (see Fig. \ref{figureFEzX}(a) without spin) and S$_y$ at k$_x$=0 (see Fig. \ref{figureFEzY}(b) without spin) is typical of the functional form of the Rashba effect, confirming that only the
Rashba spin-splitting is present beyond the relativistic
spin-momentum locking. For an electric field along the z-axis, the relativistic spin-momentum locking is composed of p$_y$, p$_x$ and d$_{yz}$  for the S$_x$, S$_y$ and S$_z$ components, respectively.
In this case, $\alpha_x\approx\alpha_y$ since the energy differences between spin-up and spin-down are of the same order of magnitude in Figs. \ref{figureFEzX}(b) and \ref{figureFEzY}(a).
On the contrary, the component S$_z$ exhibits the nodal plane as observed in Figs. \ref{figureFEzX}(c) and \ref{figureFEzY}(c), this agrees with what we have demonstrated in the previous section for the toy model.

\begin{figure*}[t!]
\centering
\includegraphics[width=6.8cm,angle=270]{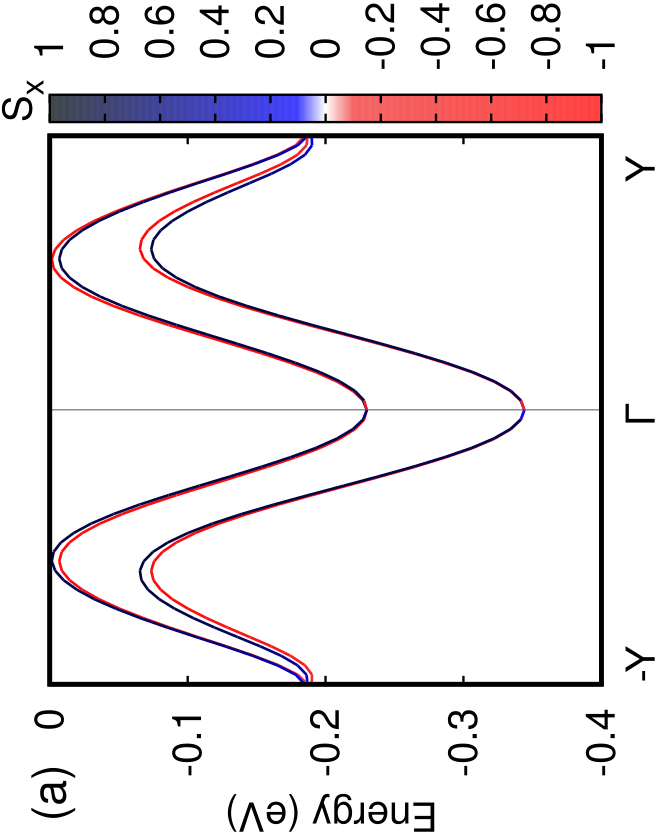}
\includegraphics[width=6.8cm,angle=270]{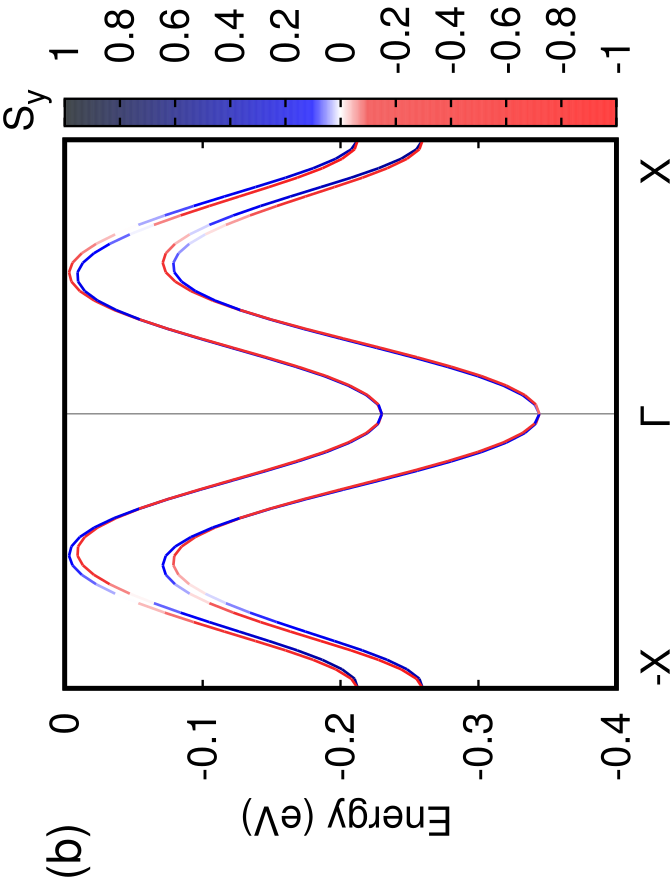}
\caption{Band structures of the A-centered phase with N\'eel vector along the $b$-axis and antiferroelectric distortions along the $y$-axis. (a) S$_x$ component along the $\overline{Y}$-$\Gamma$-Y direction. 
(b) S$_y$ component along the $\overline{X}$-$\Gamma$-X direction. These results reproduce a Rashba-type spin-orbit system with an effective electric field along the $z$-axis.
}\label{figure12}
\end{figure*}
\begin{figure*}[t!]
\centering
\includegraphics[width=7.2cm,angle=270]{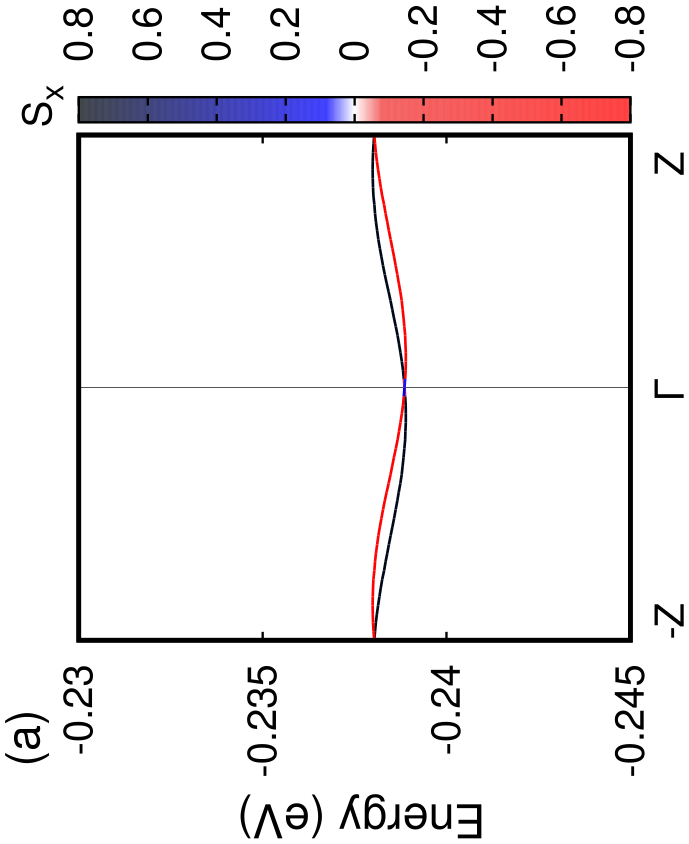}
\includegraphics[width=7.2cm,angle=270]{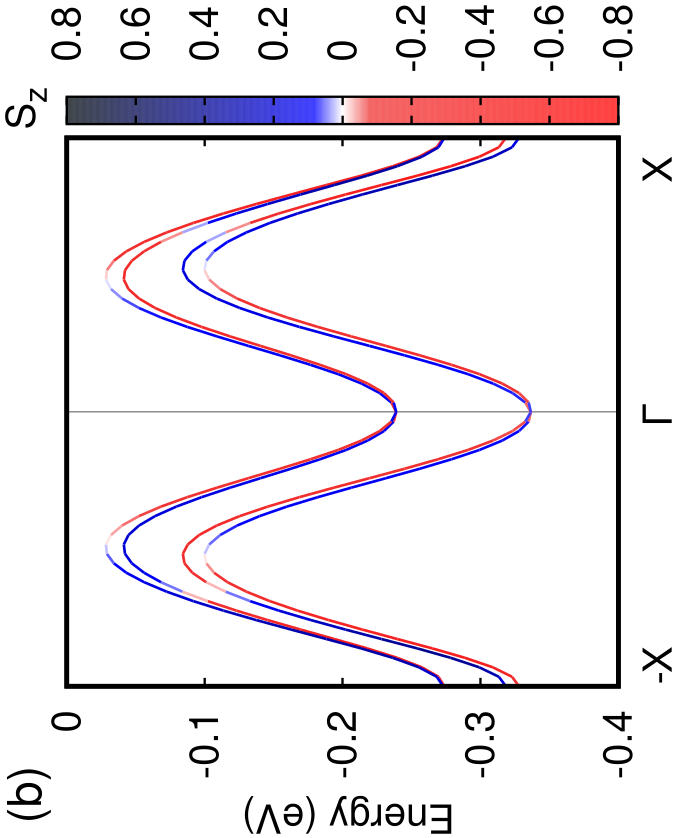}
\caption{Band structures of the A-centered phase with N\'eel vector along the $b$-axis and antiferroelectric distortions along the $z$-axis. (a) S$_x$ component along the $\overline{Z}$-$\Gamma$-Z direction. 
(b) S$_z$ component along the $\overline{X}$-$\Gamma$-X direction. These results reproduce a Rashba spin-orbit system with an effective electric field along the $y$-axis.} \label{figure13}
\end{figure*}

When the electric field is applied along the x-direction (E$_x$), the Rashba Hamiltonian reads: 
\begin{equation}
H_{R}=({\alpha_y}\sin{k_y}\sigma^{spin}_z-{\alpha_z}\sin{k_z}\sigma^{spin}_y)\sigma_0^{site}
\end{equation}
with $\alpha_y>>\alpha_z$ for the same reasons explained in the previous case. The results obtained by DFT are reported in Figs. 
\ref{figureFExY} and \ref{figureFExZ}.
Analogously to the previous case, Figs. \ref{figureFExY}(c) and \ref{figureFExZ}(b) prove the destruction of the nodal plane at kx=0 for the S$_z$ and S$_y$ components. The spin-momentum locking of S$_z$ becomes p$_y$, while the spin-momentum locking of S$_y$ becomes p$_z$. The absence of spin spectral weight of S$_y$ at k$_z$=0 (see Fig. \ref{figureFExY}(a) without spin) and S$_z$ at k$_y$=0 (see Fig. \ref{figureFExZ}(c) without spin) is typical of the functional form of the Rashba effect. 
On the contrary, the component S$_x$ exhibits the nodal plane as observed in Figs. \ref{figureFExY}(a) and \ref{figureFExZ}(a).
For an electric field along the x-axis, the relativistic spin-momentum locking is composed of d$_{yz}$, p$_z$ and p$_y$, for the S$_x$, S$_y$ and S$_z$ components, respectively.
\\

In the case of an electric field applied along the $y$ direction (E$_y$), the SOC assumes, close to the $\Gamma$ point, the Rashba form of:
\begin{equation}
    H_{R}=({\alpha_x}\sin{k_x}\sigma^{spin}_z-{\alpha_z}\sin{k_z}\sigma^{spin}_x)\sigma_0^{site}
\end{equation}
where the dominant component $S_y$ is not affected by the Rashba.
From Figure \ref{figureFEyX}(c), we can observe that k$_y$=0 is no longer a nodal plane for the S$_z$ component and the only nodal plane left is the k$_x$=0; therefore, this proves that spin-momentum locking of S$_z$ becomes p$_y$. From Figure \ref{figureFEyZ}(a), we can observe that k$_y$=0 is no longer a nodal plane for the S$_x$ component and the only nodal plane left is k$_z$=0; therefore, this proves that spin-momentum locking of S$_x$ becomes p$_z$. 
The absence of spectral weight of S$_x$ at k$_z$=0 (see Fig. \ref{figureFEyX}(a) without spin spectral weight) and S$_z$ at k$_x$=0 (see Fig. \ref{figureFEyZ}(c) without spin spectral weight) is typical of the functional form of Rashba effect, confirming that only the Rashba spin-splitting is present beyond the relativistic spin-momentum locking. We can observe a clear p-wave symmetry for S$_x$ and S$_z$ along the nodal planes deriving from the Rashba effect, which confirms that the Rashba effect destroys the d-wave spin-momentum locking of the component parallel to the electric field. 
The preservation of the nodal plane for the S$_y$ component observed in Figs. \ref{figureFEyX}(b) and \ref{figureFEyZ}(b) are a confirmation that the spin-momentum locking is preserved for the S$_y$ component. The Rashba effect, which is a relativistic p-wave spin-momentum locking, is the only one present on the nodal plane; however, in the generic k-point, the Rashba co-exists with the spin-momentum locking of the given spin component, destroying the d-wave symmetry and can lead to a p-wave symmetry if the centrosymmetric spin-momentum locking is suitable. When the electric field is along the y-axis, the relativistic spin-momentum locking is composed of p$_z$, d$_{xz}$ and p$_y$ for the S$_x$, S$_y$ and S$_z$ components, respectively. In Fig. \ref{figureFEyX}(c), the splitting between up and down is around 100 meV, while in Fig. \ref{figureFEyX}(a) the splitting is around 1 meV. These splittings are related to the $\alpha$ parameters; therefore, we can extract that $\alpha_x>>\alpha_z$ for this given case.

\begin{figure*}[t!]
\centering
\includegraphics[width=4.58cm,angle=270]{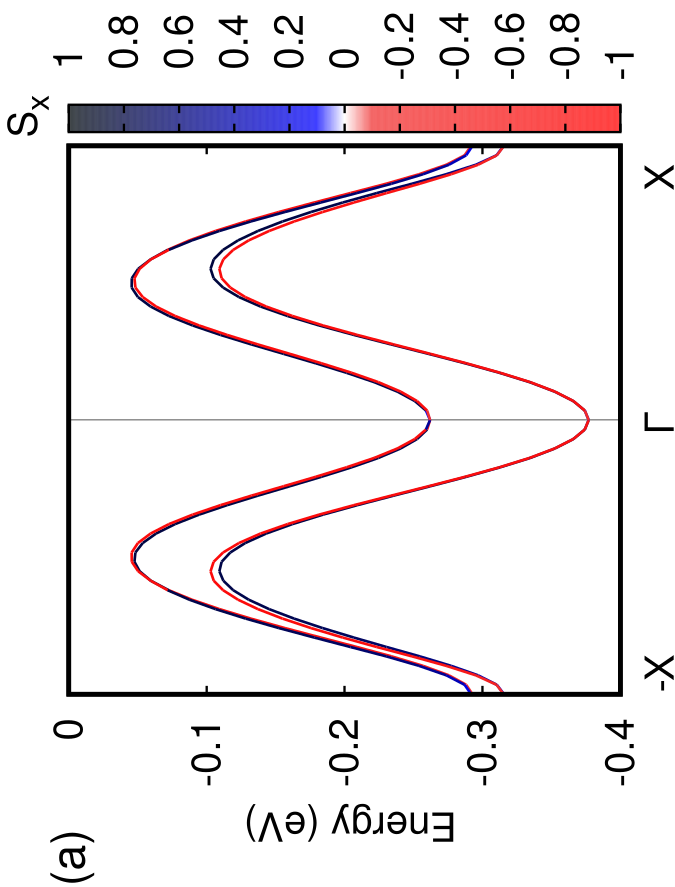}
\includegraphics[width=4.58cm,angle=270]{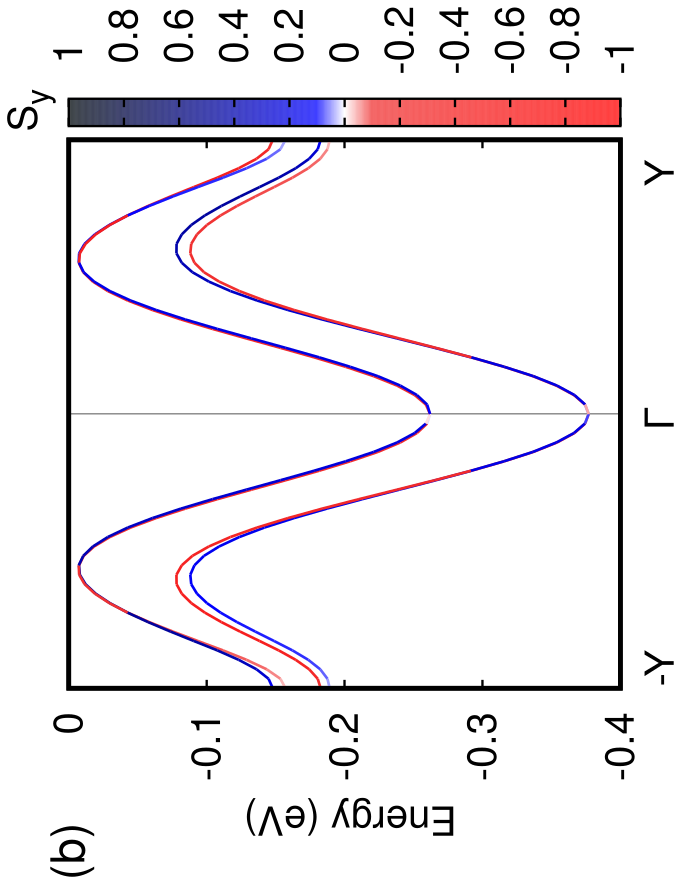}
\includegraphics[width=4.58cm,angle=270]{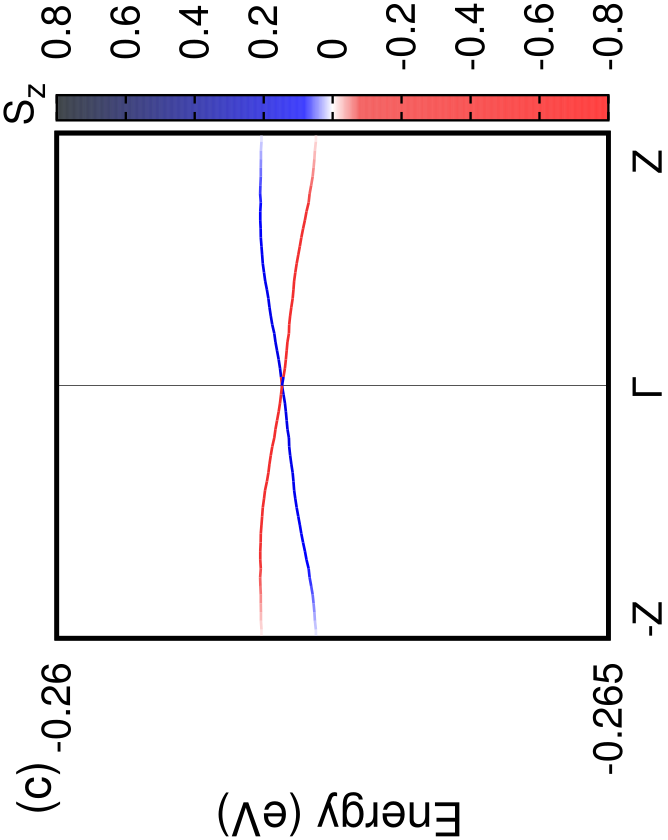}
\caption{Band structures of the A-centered phase with N\'eel vector along the $b$-axis and antiferroelectric distortions along the $x$-axis. (a) S$_x$ component along the $\overline{X}$-$\Gamma$-X direction. 
(b) S$_y$ component along the $\overline{Y}$-$\Gamma$-Y direction.
(c) S$_z$ component along the $\overline{Z}$-$\Gamma$-Z direction. These results reproduce a three-dimensional Weyl spin-orbit system\cite{Roy2022}.}
\label{figure14}
\end{figure*}

\subsection{Band structure with spin-orbit coupling and antiferroelectric distortions}


In the case of antiferroelectric distortions along the $y$-axis, our DFT results suggest the SOC assumes the Rashba form of:
\begin{equation}\label{RashbaEz}
H_{R}=({\alpha_y}\sin{k_y}\sigma^{spin}_x-{\alpha_x}\sin{k_x}\sigma^{spin}_y)\sigma_0^{site}
\end{equation}
which is equivalent to an effective electric field applied along the $z$-axis. 
The DFT results for this case are reported in Fig. \ref{figure12}, where we have included only the band structures where we can observe the breaking of the nodal planes.
The reason for this effective electric field along the z-direction can be related to the Born effective charge. Indeed, it was demonstrated that the Born effective charge Z$_{ij}$ tensor is not diagonal for Ca$_2$RuO$_4$\cite{ma15196657}, therefore, the displacement of Ru atoms in opposite directions along the $y$-axis can cancel each other for the y-direction and generate an effective electric field along the $z$-axis responsible for the Rashba effect reported in equation (\ref{RashbaEz}). 
In the case of antiferroelectric distortions along the $z$-axis, the SOC leads to the Rashba form:
\begin{equation}\label{RashbaEy}
H_{R}=({\alpha_z}\sin{k_z}\sigma^{spin}_x-{\alpha_x}\sin{k_x}\sigma^{spin}_z)\sigma_0^{site}
\end{equation}
which is equivalent to an effective electric field applied along the $y$-axis. The DFT results for this case are reported in Fig. \ref{figure13}, where we have included only the band structures showing the breaking of the nodal planes. In the antiferroelectric case for displacement along the y and z-axis, the Rashba spin splittings between spin-up and spin-down are much smaller than in the ferroelectric cases. This means that the values of the Rashba couplings $\alpha_x$, $\alpha_y$ and $\alpha_z$ are much smaller than the corresponding values in the ferroelectric cases. Another consequence of the small values of the Rashba couplings is the sign change of the spin polarization across the bands observed in \ref{figure12}(b) and \ref{figure13}(b) with a color change. A stronger Rashba coupling will produce fully polarized bands with the same spin from -X to X, as we confirmed by our model Hamiltonian.

In the case of an antiferroelectric distortion along the $x$-axis, the Hamiltonian for the SOC assumes the Weyl form that reads:
\begin{equation}
    H_{W}=({\alpha_x}\sin{k_x}\sigma^{spin}_x+{\alpha_y}\sin{k_y}\sigma^{spin}_y+ \\{\alpha_z}\sin{k_z}\sigma^{spin}_z)\sigma_0^{site}
\end{equation}
The DFT band structures are shown in Fig. \ref{figure14}, where we have included only the band structure showing the breaking of the nodal plane of the spin-momentum locking. We can observe that in this case, the spin-momentum locking is broken for all three spin components. From the spin-splittings, we infer that the $\alpha$ constants are of the same order of magnitude as the other antiferroelectric distortions.
This Weyl-type SOC arises in space group no. 19, unlike the Rashba-type, which occurs in space group no. 29. The Weyl-type spin-orbit (as the Rashba) does not induce any weak ferromagnetism; at the same time, the Weyl-type spin-orbit breaks the spin-momentum locking for all three spin components but preserves the vanishing magnetization. We can examine this in more detail by focusing on the S$_x$ component (since the other behaves similarly), where we have the d$_{yz}$ from the relativistic spin-momentum locking of the centrosymmetric phase and p$_{x}$ from the Weyl SOC. In this case, the spin-momentum locking exhibits no nodal plane; consequently, the interplay between the spin-momentum locking induced by altermagnetism and that arising from the Weyl SOC precludes the formation of any p-wave magnetic order. This can also be expressed in terms of magnetic dipole and magnetic quadrupole in the k-space\cite{AutieriRSML,Cuono26}, saying that the presence of the p-wave depends on the interplay between the geometrical properties of the quadrupole and dipole. Although the Weyl form breaks both the k$_y$=0 and k$_z$=0 nodal planes, the system nevertheless retains symmetry-protected nodal lines (NLs) coincident with the k$_y$ and k$_z$ axes. 
\\

The summary of the interplay between relativistic spin-momentum locking and the Rashba or Weyl SOC is reported in Table \ref{tab:RSML}. The relevant ingredients are the space group and the kind of breaking of inversion symmetry.
It is important to note that, in addition to SOC splitting, the relativistic spin–momentum locking of the centrosymmetric phase plays a crucial role; depending on its nature, the system may or may not exhibit p-wave components in the spin–momentum locking or nodal lines. In simple terms, whenever we consider a combination of a $d$-wave and a $p$-wave for a given spin component, the resulting spin-momentum locking can be classified into two distinct cases.
In the first case, the combination of $d_{xy}$ and $p_x$ leads to a band structure in which only a nodal plane at $k_x = 0$ appears. To first order, the spin-momentum locking can therefore be approximated by a $p_x$ component.
In the second case, the combination of $d_{xy}$ and $p_z$ behaves differently. Since there is no momentum plane where both $d_{xy}$ and $p_z$ simultaneously vanish, nodal planes do not occur. Instead, the system exhibits only nodal lines.

\begin{table}[h!]
\centering
\begin{tabular}{|c|c|c|c|c|}
\hline
Symmetry (space group) & Rashba/Weyl & S$_x$ & S$_y$ & S$_z$ \\ \hline
non-relativistic CS (no. 61) & - & - & d$_{xz}$ & - \\ \hline
relativistic CS (no. 61) & - & d$_{yz}$ & d$_{xz}$ & d$_{xy}$\\ \hline
FEx (no. 29) & Rashba with E$_x$ & d$_{yz}$ & p$_z$ & p$_y$ \\ \hline
FEy (no. 29) & Rashba with E$_y$ & p$_z$ & d$_{xz}$ & p$_x$ \\ \hline
FEz (no. 29) & Rashba with E$_z$ & p$_y$ & p$_x$ & d$_{xy}$ \\ \hline
AFEx (no. 19) & Weyl & NLs & NLs & NLs \\ \hline
AFEy (no. 29)& Rashba with E$_z$ & p$_y$ & p$_x$ & d$_{xy}$ \\ \hline
AFEz (no. 29)& Rashba with E$_y$ & p$_z$ & d$_{xz}$ & p$_x$ \\ \hline
\end{tabular}
\caption{Summary of the spin-momentum locking for centrosymmetric, ferroelectric- and antiferroelectric-like distortions. In the relativistic CS case, the spin-orbit
produces spin cantings, but no Rashba or Weyl spin splittings are
present.}
\label{tab:RSML}
\end{table}

\section{The case of the stripes as modulated electric field}

Finally, we investigated a modulated electric field introducing atomic shifts of a single layer of Ru in the Ca$_2$RuO$_4$ unit cell as reported in Fig. \ref{Displacements}(c). The crystal structure with this kind of distortion is symmetrically equivalent to the interface between different domains as observed in the stripes (or ferroelastic domains) of Ca$_2$RuO$_4$ under electric field quenching \cite{Gauquelin23}. This section is divided into three subsections: in the first, we demonstrate what happens in the presence of two non-relativistic spin–momentum lockings; in the second, we confirm these results in our system from the density functional theory band structure; and in the third, we show that this case always exhibits weak ferromagnetism. \\

\begin{figure}[t!]
\centering
\includegraphics[width=7.99cm,angle=0]{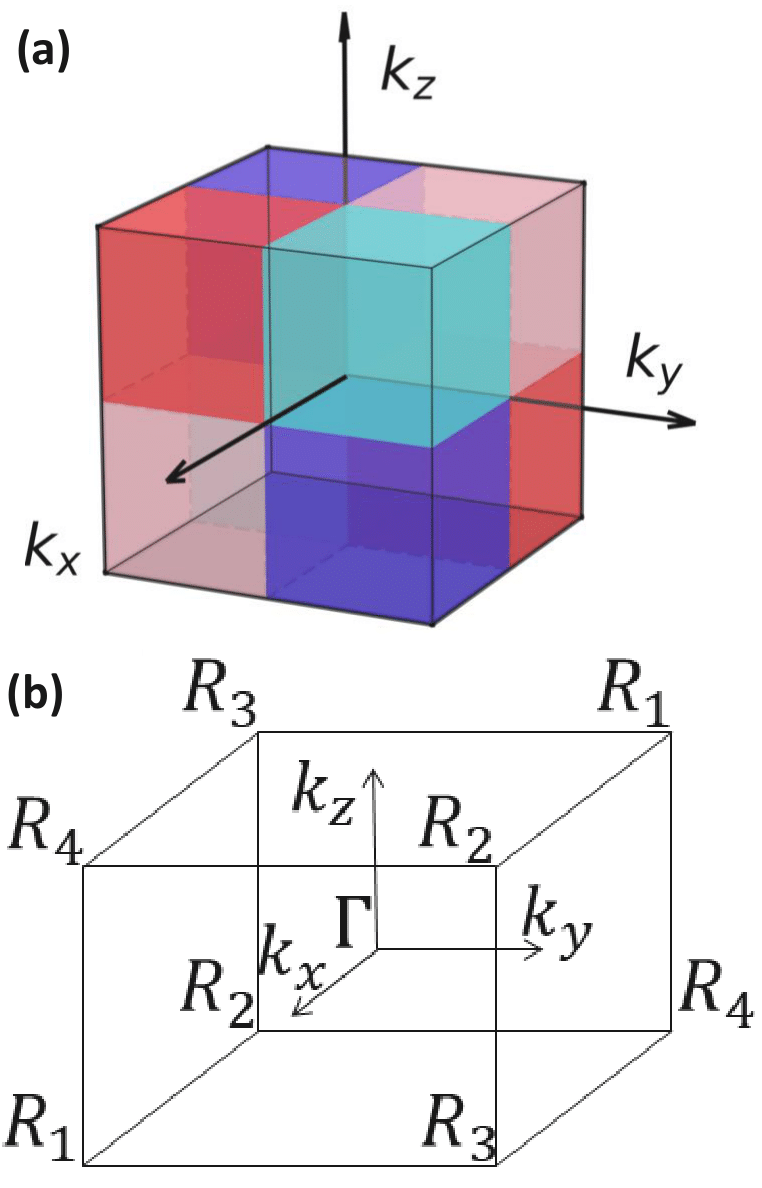}
\caption{(a) Non-relativistic spin-momentum locking and (b) Brillouin zone of the Ca$_2$RuO$_4$ in the presence of an interface or modulated electric field along the $a$-, $b$- or $c$-direction for the A-centered magnetism. Dark red and dark blue regions have opposite non-relativistic spin-splitting. Light blue and light red regions have opposite non-relativistic spin-splitting.
In our notation, the high-symmetry points with the subscripts 1, 2, 3 and 4 show altermagnetism along the path towards the $\Gamma$ point. The lines $\Gamma$-R$_1$ and $\Gamma$-R$_2$ show opposite spin-splitting. The same happens for $\Gamma$-R$_3$ and $\Gamma$-R$_4$. The nodal plane appears only for k$_x$=0. As a consequence, we have altermagnetic surface states on two of the three main surfaces (010) and (001). This represents a system with two distinct spin-momentum lockings.}\label{BZ_modulation}
\end{figure}

\subsection{Two Non-relativistic Spin-momentum lockings}

Starting from the non-relativistic spin–momentum locking observed in our system, we aim to show with an analytical model that this behavior can be described with two distinct types of non-relativistic spin–momentum locking.
In Fig. \ref{BZ_modulation}, we show the position of the high-symmetry points of the spin-momentum locking in the Brillouin zone. Unlike the centrosymmetric case and the cases with ferroelectric or antiferroelectric distortions, our system exhibits four inequivalent regions of the spin-momentum locking as represented in Fig. \ref{BZ_modulation}(a) with different colors and four inequivalent R points listed in Fig.~\ref{BZ_modulation}(b), and named as R$_1$, R$_2$, R$_3$ and R$_4$.
This kind of non-relativistic spin-momentum locking can be described by the linear combination of two distinct altermagnetic planar d-wave orders; these two altermagnetic orders are the planar d$_{xz}$ of the Ca$_2$RuO$_4$ bulk\cite{Cuono23orbital} and the planar d$_{xy}$ of the twodimensional Ca$_2$RuO$_4$ \cite{D4NR04053H}. 
The d$_{xy}$ altermagnetism in Ca$_2$RuO$_4$ is an example of hidden altermagnetism\cite{hiddenaltermagnetism,mazin2023induced}, where a lowering of the crystal symmetry is needed to observe it.
For simplicity and intuitive notation, we define these non-relativistic spin-momentum lockings and their associated hopping parameters as 3D and 2D. 
\begin{figure}[t!]
\centering
\includegraphics[width=6.19cm,angle=270]{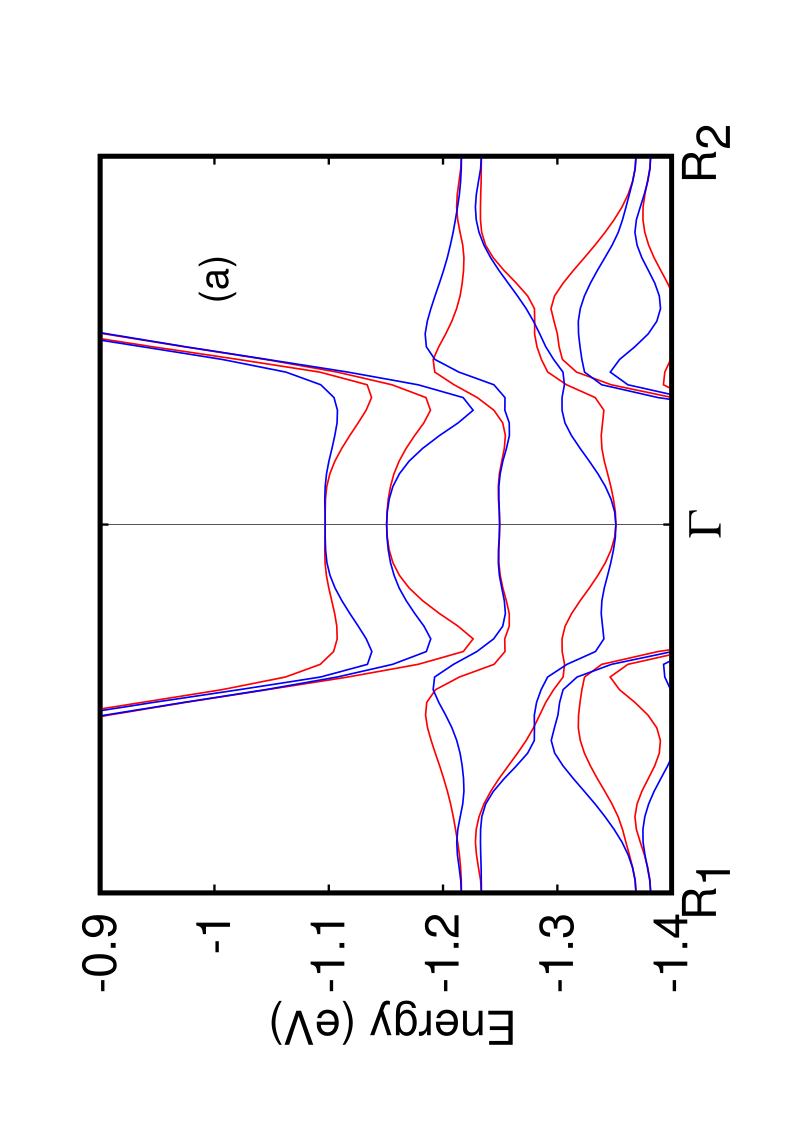}
\includegraphics[width=6.19cm,angle=270]{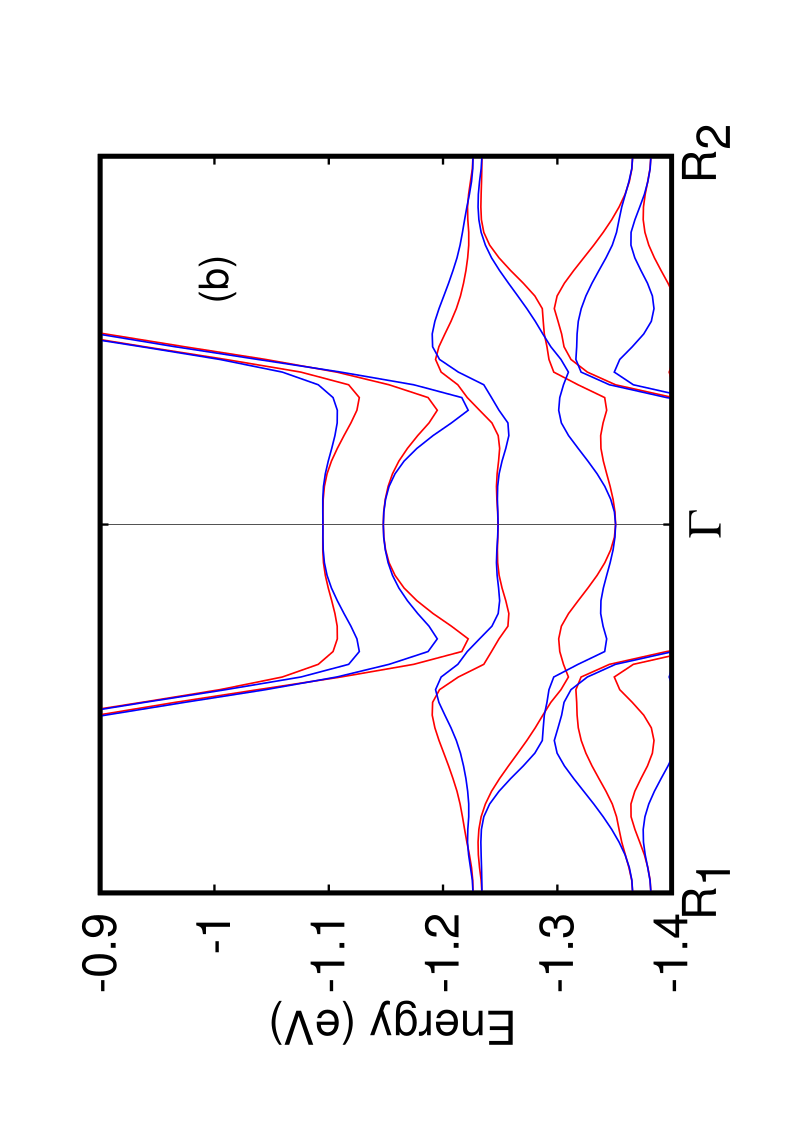}
\includegraphics[width=6.19cm,angle=270]{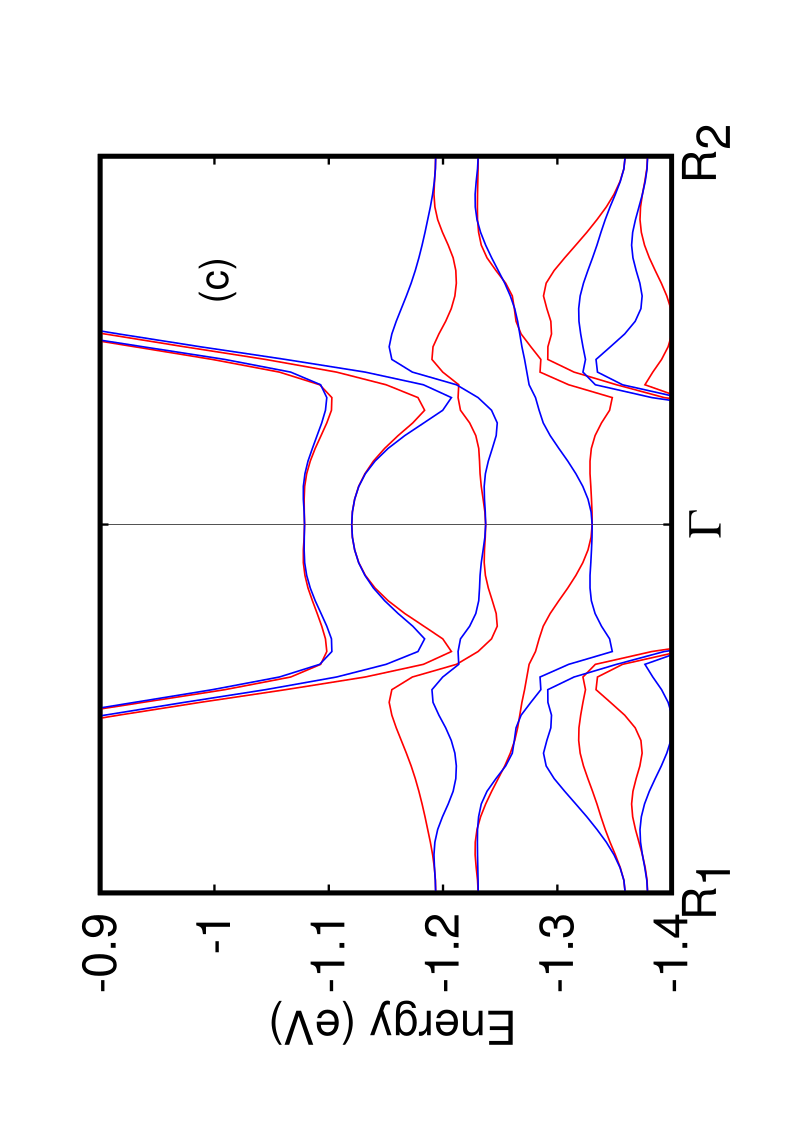}
\caption{Zoom of the band structure of Ca$_2$RuO$_4$ along the high-symmetry positions R$_1$-$\Gamma$-R$_2$ in the antiferromagnetic A-centered configuration and in the
presence of an interface or modulated electric field for different layers with Ru shifted along: (a) the $x$-axis, (b) the $y$-axis,
(c) and the $z$-axis.}\label{BS_CAO_modulation_NO_SOC_zoom}
\end{figure}
While the minimal model is extremely complex and reported in the supplementary materials, we propose the following simplified single-orbital non-relativistic Hamiltonian for the spin-up and spin-down sectors:
\begin{align} 
\nonumber
\mathcal{H}^{\rm AM}_{Sz}  = & {\Delta_z}\sigma_z^{spin}\sigma_z^{site}+ \\ 
\label{model2SML} & (4t_{am}^{3D}\sin{k_x}\sin{k_z} + 4t_{am}^{2D}\sin{k_x}\sin{k_y})\sigma_0^{spin}\sigma_z^{site}  
\end{align}
where $t_{am}^{3D}$ and $t_{am}^{2D}$ are the hopping parameters driving the two altermagnetic spin-momentum lockings, $\Delta_z$ is the on-site spin-splitting and H$^{\uparrow\downarrow}$=H$^{\downarrow\uparrow}$=0 as shown in the matrix form in the supplementary materials. The eigenvalues from $\Gamma$ to R$_2$ can be obtained by parameterizing the k-path as $\textbf{k}$=(t,t,t) with the parameter $t$ ranging from 0 to $\pi$.
If we consider only the majority spins (which are the ones containing 
$-\Delta_z$, the minority will be the same but with $+\Delta_z$), we obtain the eigenvalues:
\begin{equation}\label{path1}
    \varepsilon_{\pm}(t)=\Delta_z \pm 4(t_{am}^{3D}+t_{am}^{2D})\sin{t}^2
\end{equation}
and the same eigenvalues with inverted spin-up and spin-down are obtained for the k-path between $\Gamma$ to R$_1$. 
The eigenvalues from $\Gamma$ to R$_3$ can be obtained by parameterizing the k-path as $\textbf{k}$=(t,-t,t) with the parameter $t$ ranging from 0 to $\pi$.
If we consider only the majority spins, we obtain the eigenvalues:
\begin{equation}\label{path2}
    \varepsilon_{\pm}(t)=\Delta_z \pm 4(t_{am}^{3D}-t_{am}^{2D})\sin{t}^2
\end{equation}
and the same is obtained for the path between $\Gamma$ and R$_4$ with inverted spin-up and spin-down. 
In equation (\ref{path1}), the non-relativistic spin-splitting is driven by an effective hopping equal to $(t_{am}^{3D}+t_{am}^{2D})$, while in equation (\ref{path2}), the non-relativistic spin-splitting is driven by an effective hopping equal to $(t_{am}^{3D}-t_{am}^{2D})$. 
What we obtain is still an altermagnetic system, since the magnetization is zero in the non-relativistic limit, protected by rotational symmetries, and the system breaks time-reversal symmetry. Different from the altermagnetic system with one spin-momentum locking, where there are an even number of nodal planes, in this system, there is only one nodal plane, which is k$_x$=0. As a result, the (100) surface, which corresponds to $k_x = 0$, does not host altermagnetic surface states. Instead, the altermagnetic surface states emerge on the other two principal surfaces: (010) and (001). This model with two distinct spin-momentum lockings, described by equation (\ref{model2SML}), was confirmed by DFT calculations performed along the planes k$_x$=0, k$_y$=0 and k$_z$=0. This behavior with two spin–momentum lockings is likely only possible in systems with low crystal symmetry containing four or more magnetic atoms, with pairs of magnetic atoms exhibiting inequivalent behavior.

\begin{figure}[t!]
\centering
\includegraphics[width=6.19cm,angle=270]{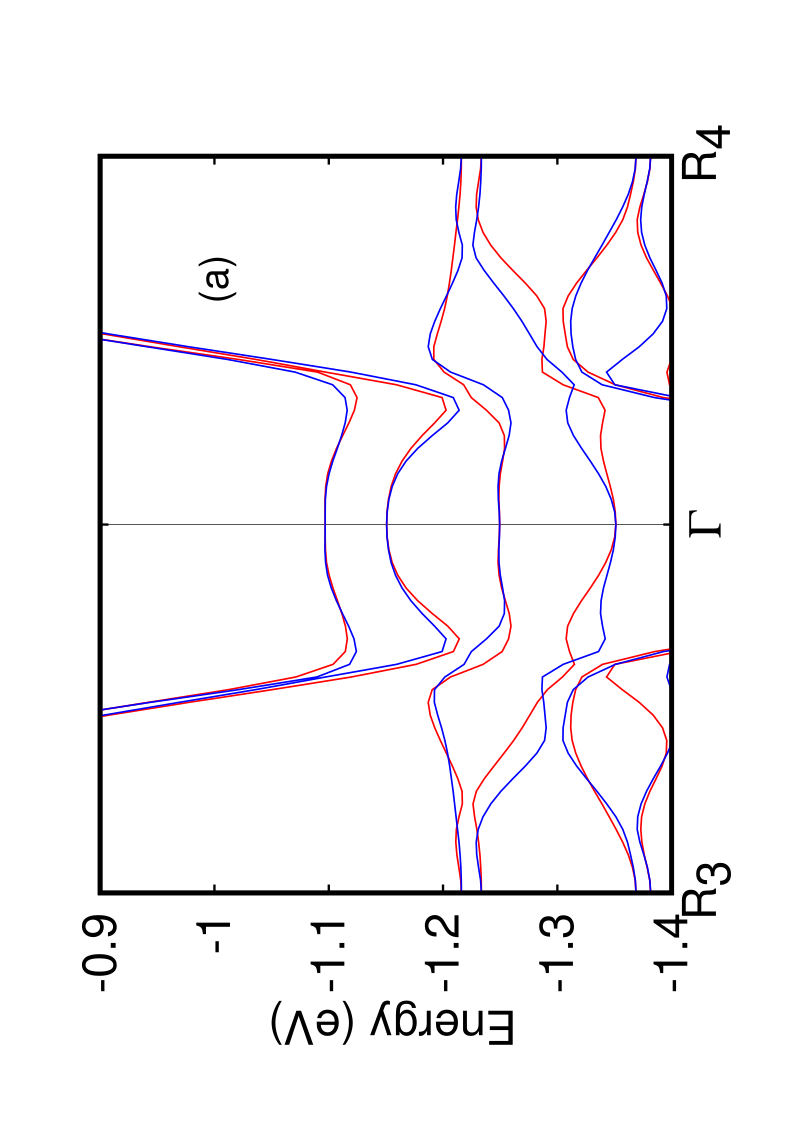}
\includegraphics[width=6.19cm,angle=270]{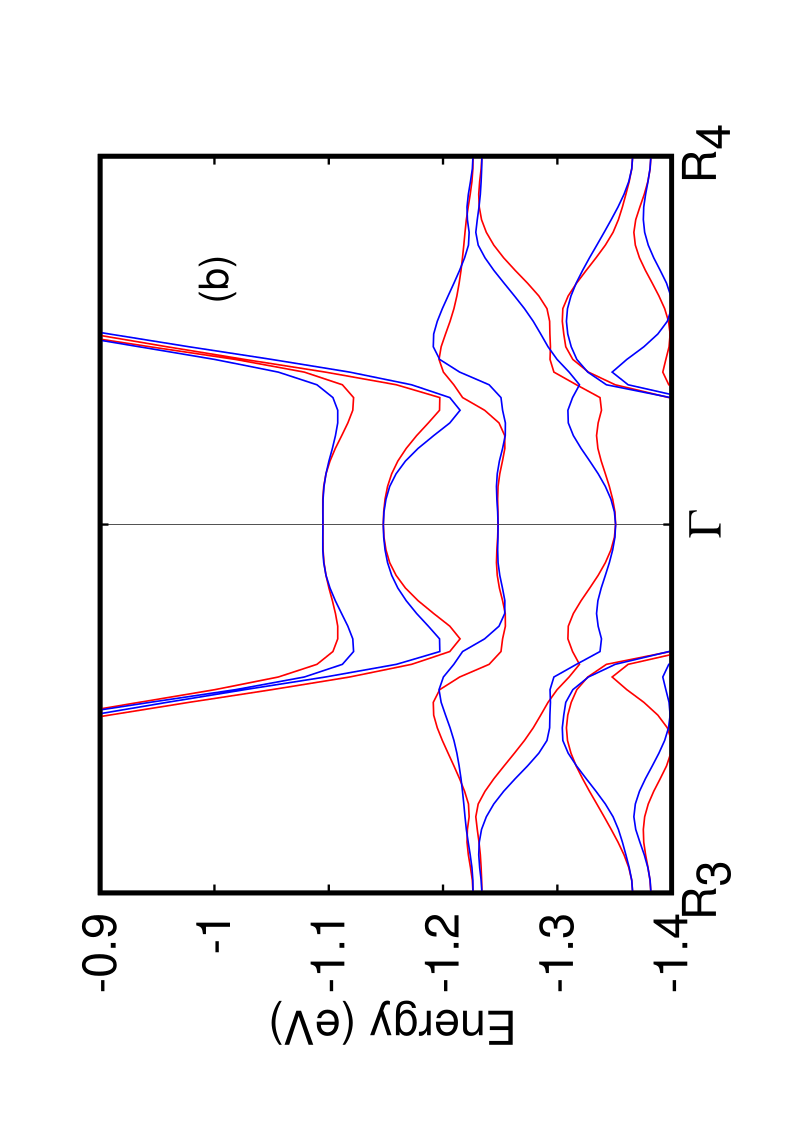}
\includegraphics[width=6.19cm,angle=270]{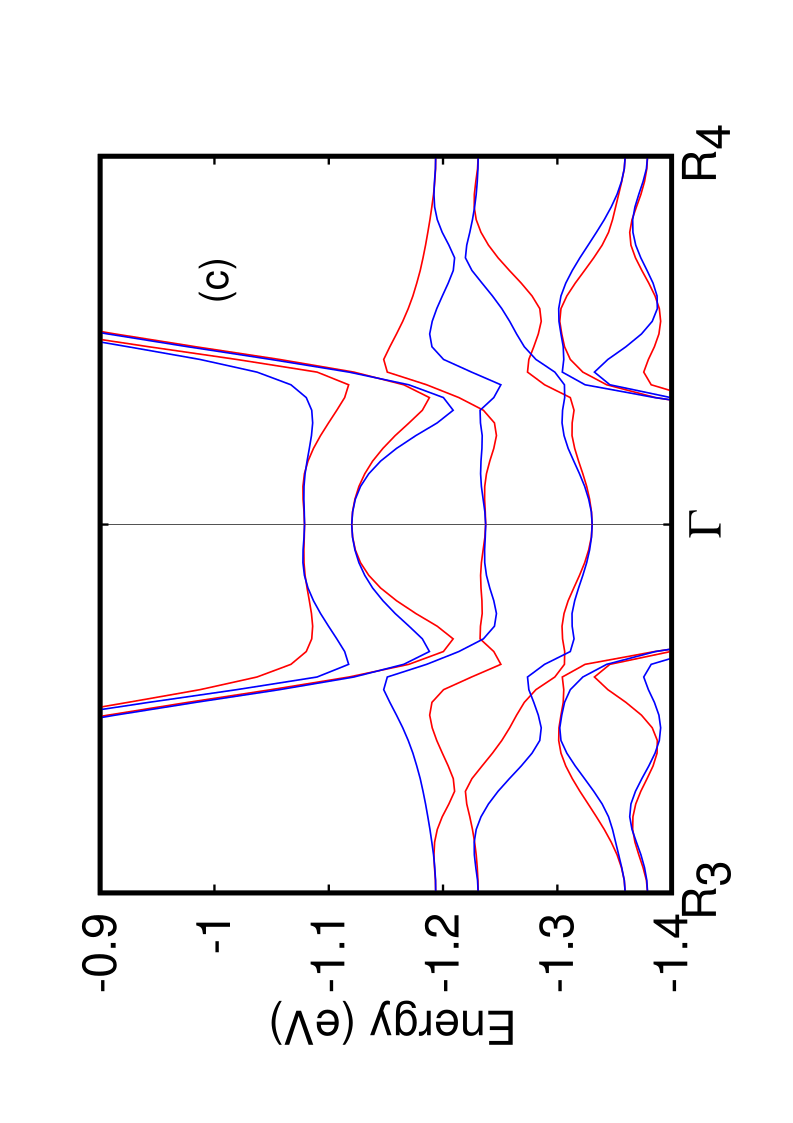}
\caption{Zoom of the band structure of Ca$_2$RuO$_4$ along the high-symmetry positions R$_3$-$\Gamma$-R$_4$ in the antiferromagnetic A-centered configuration and in the
presence of an interface or modulated electric field for different layers with Ru shifted along: (a) the $x$-axis, (b) the $y$-axis,
(c) and the $z$-axis.}\label{BS_CAO_modulation_NO_SOC_zoom_R3_R4}
\end{figure}


\subsection{DFT band structures for two non-relativistic spin-splittings}

The band structure exhibits opposite non-relativistic spin-splitting along the lines $\Gamma$-R$_1$ and $\Gamma$-R$_2$, the same occurs for $\Gamma$-R$_3$ and $\Gamma$-R$_4$. However, the bands along the  R$_1$-$\Gamma$-R$_2$ k-path differ from those along R$_3$-$\Gamma$-R$_4$. 
Figs. \ref{BS_CAO_modulation_NO_SOC_zoom} and \ref{BS_CAO_modulation_NO_SOC_zoom_R3_R4} show zoomed views of the band structure along R$_1$-$\Gamma$-R$_2$ and R$_3$-$\Gamma$-R$_4$, respectively. In Figs.~\ref{BS_CAO_modulation_NO_SOC_zoom}(a), (b) and (c), we report the band structure along one k-path for displacements of Ru atoms along the x-, y-, and z-directions, respectively, while Figs.~\ref{BS_CAO_modulation_NO_SOC_zoom}(a), (b) and (c) present the corresponding results along the other k-path.
We observe minor changes in the band structure among the different shift directions, but overall, the spin-moment locking and the size of the spin-splitting are consistent for all shift directions.

As previously shown \cite{Cuono23orbital}, the band structures of Ca$_2$RuO$_4$ consist of two groups of bands. The first group includes four weakly dispersive bands located around -1 eV and four weakly dispersive bands near +1 eV to the Fermi level. The second group is composed of the more dispersive bands, comprising two bands at the top of the valence region and two additional bands with a bandwidth ranging from -0.5 to -2.0 eV. The weakly dispersive bands near -1 eV and +1 eV predominantly exhibit ${{\gamma}z}$ orbital character, while the more dispersive bands are mainly of ${xy}$ character. While in the centrosymmetric case, the non-relativistic spin-splitting is absent in the ${xy}$ subsector, in the presence of an interface or a modulated electric field, the non-relativistic spin-splitting also appears in the ${xy}$ bands. 

\begin{table*}[!] 
\centering
\begin{tabular}{|c|c|c|c|c|c|c|c|c|c|c|c|c|}
\hline
Distortion & M$_1$ & M$_2$ & M$_3$ & M$_4$ & M$_{\text{tot}}$ & m$_a$ & m$_b$ & m$_c$ & $E_{G}$\\
\hline
 $\Vec{n}$ $\parallel$ $\Vec{a}$, Ru shift along x & (m$_a$,m$_b$,m$_c$)   & (-m'$_a$,-m'$_b$,m'$_c$)  & (m'$_a$,-m'$_b$,m'$_c$) & (-m$_a$,m$_b$,m$_c$) & (0,M$_y$,4m$_c$) & 1.45 & 0.02 & -0.02  & 0.85\\
\hline
 $\Vec{n}$ $\parallel$ $\Vec{a}$, Ru shift along y& (m$_a$,m$_b$,m$_c$)  & (-m'$_a$,-m'$_b$,m'$_c$)  & (m'$_a$,-m'$_b$,m'$_c$)  & (-m$_a$,m$_b$,m$_c$)  & (0,M$_y$,4m$_c$) & 1.45 & 0.02 & -0.04  & 0.85\\
\hline
$\Vec{n}$ $\parallel$  $\Vec{a}$, Ru shift along z &    (m$_a$,m$_b$,m$_c$) & (-m'$_a$,m'$_b$,m'$_c$)  & (m'$_a$,m'$_b$,m'$_c$) & (-m$_a$,m$_b$,m$_c$) &  (0,M$_y$,4m$_c$) & 1.46  & -0.01 & -0.04 & 0.88 \\
\hline
$\Vec{n}$ $\parallel$  $\Vec{b}$, Ru shift along x & (m$_a$,m$_b$,m$_c$)   & (-m'$_a$,-m'$_b$,m'$_c$)  & (-m'$_a$,m'$_b$,-m'$_c$) & (m$_a$,-m$_b$,-m$_c$) & (M$_x$,0,0) & -0.01 & 1.45 & -0.12 & 0.85  \\
\hline
$\Vec{n}$ $\parallel$  $\Vec{b}$, Ru shift along y & (m$_a$,m$_b$,m$_c$)  & (-m'$_a$,-m'$_b$,m'$_c$)  & (-m'$_a$,m'$_b$,-m'$_c$)& (m$_a$,-m$_b$,-m$_c$) & (M$_x$,0,0)  & -0.01  & 1.45& -0.11  & 0.87 \\
\hline
 $\Vec{n}$ $\parallel$ $\Vec{b}$, Ru shift  along z   & (m$_a$,m$_b$,m$_c$)  &  (-m'$_a$,-m'$_b$,m'$_c$)& (-m'$_a$,m'$_b$,-m'$_c$) & (m$_a$,-m$_b$,-m$_c$) & (M$_x$,0,0) & 0.01 & 1.45 & -0.06 & 0.88 \\
\hline
\end{tabular}
\caption{Magnetic moments and their symmetries, total magnetization and band gap at the $\Gamma$ point of the different magnetic configurations studied only for the A-centered phase with N\'eel vector ($\Vec{n}$) along $\Vec{a}$ and $\Vec{b}$ with displacements of the Ru atoms only in one layer along x,y, and z. m$_a$, m$_b$, m$_c$ are reported in $\mu_B$ and they represent the magnetizations of the Ru atoms located at the coordinate z=0. m'$_a$, m'$_b$ and m'$_c$ are the magnetizations at z=0.5, these values defer slightly by a factor of $10^{-2}$.  $E_{G}$ is reported in eV. M$_x$ and M$_y$ represent a small magnetization rising from the magnetizations in the two inequivalent layers at z=0 and z=0.5.}
\label{tab:label_shift}
\end{table*}

\subsection{Weak ferromagnetism and gap reduction in the stripe phase}

When we shift the Ru atoms in just one layer to simulate the stripes, we further lower the crystal symmetries with respect to the ferroelectric and antiferroelectric distortions. In the case of the stripes, for a shift along the $x$-axis we have P2$_1$ (space group no. 4), while along the $y$- and $z$-axis we have space group Pc (space group no. 7). The dominant spin component is 1.45 $\mu_B$, which is basically unchanged with respect to the centrosymmetric case as reported in Table \ref{tab:label_shift}.
The complete DFT results for the magnetic moments for the ground state (A-centered with N\'eel vector along the $b$-axis) and the second lowest energetic configuration (A-centered with N\'eel vector along the $a$-axis) are also reported.
In both magnetic configurations, the system exhibits an additional weak ferromagnetism as reported in Table \ref{tab:label_shift}. Indeed, due to the asymmetry between the different layers of Ru atoms, we have different sets of magnetic moments which we define with m and m' as reported in Table \ref{tab:label_shift}. For the ground state, we have weak ferromagnetism along the $x$-axis named M$_x$, while for the second lowest energetic configuration, we have an additional weak ferromagnetism named M$_y$ along the $y$-axis, which co-exists with the previous weak ferromagnetism along the $c$-axis. M$_x$ and M$_y$ represent a small magnetization that is smaller than the average canting values in this compound.
The origin of M$_x$ and M$_y$ should be likely attributed to the hidden spin-momentum locking 4t$_{am}^{2D}\sin{(k_x)}\sin{(k_y)}$.

For the centrosymmetric case with A-centered magnetism and $\Vec{n}$ along the $b$-axis, the band gap is 0.89 eV. In the case of a modulated electric field, the band gap gets reduced in the range of 0.85-0.88 eV as reported in Table \ref{tab:label_shift}. Therefore, in the stripe case with spin-orbit, there is a gap reduction due to the weak ferromagnetism and a gap reduction due to the breaking of the Kramer's degeneracy in the ${xy}$ (top of the valence band). These two effects combine to produce a reduction of the band gap in the stripe case.\\

\section{Discussion and Conclusions}

We study the interplay between relativistic spin-momentum locking and the breaking of the inversion symmetry. We develop new strategies to address, by using first-principle calculations, the coexistence of multiple spin-momentum lockings and Rashba-Weyl SOC.

As a testbed, we use the quasi-two-dimensional Ca$_2$RuO$_4$, which hosts relativistic spin-canting and several degrees of freedom. First, we studied the centrosymmetric case as a benchmark and then we moved to non-centrosymmetric cases. Depending on the specific symmetry breaking, this can give rise to Rashba-type SOC, Weyl-type SOC, or the coexistence of two distinct spin-momentum lockings. 
Centrosymmetric bulk Ca$_2$RuO$_4$ hosts as non-relativistic planar d$_{xz}$ spin-momentum locking. The planes k$_x$=0 and k$_z$=0 are nodal planes where altermagnetism is absent. While the non-relativistic d$_{xz}$-altermagnetism could be described with a toy-model interband hybridization term like $\sigma_z\sin{k_x}\sin{k_y}$, our tight-binding model shows that the real system is much more complex and includes inter- and intraband hybridizations. Taken individually, a group of hybridization generates d$_{xy}$-altermagnetism while another group generates g-wave altermagnetism; however, their combination produces d$_{xz}$-altermagnetism. Ca$_2$RuO$_4$ is an orbital-selective altermagnet with altermagnetism suppressed in the $xy$ bands when in the A-centered magnetic order. However, in the B-centered magnetic order, the non-relativistic spin-splitting altermagnetism also rises in the ${xy}$ band. We conclude that orbital-selective altermagnetism is due to an accidental suppression of the influence of the altermagnetic hopping on the ${xy}$-orbitals in the B-centered magnetic order. Beyond the spin-momentum locking, in the centrosymmetric case, we also analyze the band gap and weak ferromagnetism for different magnetic orders and N\'eel vector orientations. The properties of the ground state determined theoretically, such as the magnetic order and the magnitude of the spin canting, are in good agreement with the experimental results. The easy axis is along the $b$-axis and the magnetocrystalline anisotropy is 9.4 meV per formula unit.
In the A-centered phase with Neel vector along the b-axis, Ca$_2$RuO$_4$ is a pure altermagnet, while it becomes a weak ferromagnet in the other magnetic configurations close in energy. Therefore, this is a transition from the pure altermagnet to a weak ferromagnet.

Upon including SOC, Ca$_2$RuO$_4$ exhibits a relativistic spin-canting and the relativistic spin-momentum locking. The relativistic spin-momentum locking for Ca$_2$RuO$_4$ with N\'eel vector along the $b$-axis is composed of the d$_{yz}$-wave, d$_{xz}$-wave and d$_{xy}$-wave for the S$_x$, S$_y$ and S$_z$ components, respectively.
In case of breaking of inversion symmetry by ferroelectric and antiferroelectric distortions, we observe a reduction of the band gap, the breaking of the orbital-selective altermagnetism, and above all, the rise of the Rashba-type or Weyl-type SOC; both types of SOC preserve the zero net magnetization.
In the ferroelectric cases and two antiferroelectric cases, we have the rise of an effective Rashba Hamiltonian with antisymmetric spin-texture in the nodal planes of k-space. In one antiferroelectric case, the system does not host the Rashba-type but the Weyl-type SOC.
Even if under ferroelectric- and antiferroelectric-like distortions, there are no qualitative changes in the symmetries of the non-relativistic spin-momentum locking and in the weak ferromagnetism, the relativistic spin-momentum locking of two spin components gets destroyed by the Rashba spin-splitting. We find that the spin–momentum locking of the spin component parallel to the electric field is the only one that survives in the presence of Rashba coupling. In contrast, Weyl-type spin–orbit coupling disrupts the spin–momentum locking for all three spin components. Remarkably, both types of spin-orbit preserve the vanishing magnetization since they do not induce any kind of ferromagnetism.
These results advance the understanding of altermagnetic surface states and interface states. Since the material's surfaces and interfaces host an electric field normal to the plane, only the perpendicular component sustains the spin–momentum locking; this result is general and it does not depend on the relativistic spin-momentum locking of the material under investigation.
In Ca$_2$RuO$_4$, we observed the rise of p-wave components when there is Rashba SOC and nodal lines when there is Weyl SOC; these results are not general since they critically depend on relativistic spin-momentum locking of the specific material.  
Finally, to model the stripe pattern, we introduced a modulated electric field that induces atomic displacements within a single layer of the Ca$_2$RuO$_4$ unit cell. This modulation drives a magnetic phase transition into an exotic altermagnetic state characterized by two distinct non-relativistic spin–momentum locking channels, accompanied by weak ferromagnetism. These two non-relativistic spin–momentum lockings correspond to bulk altermagnetism appearing in the three-dimensional case, while the second represents a hidden altermagnetism emerging at lower crystal symmetry or in the two-dimensional case.

There are growing expectations regarding the realization of the non-relativistic p-wave magnet\cite{Chakraborty2025,zk69-k6b2,fukaya2025tunnelingconductancesuperconductingjunctions} and p-wave magnetism from spin-spirals\cite{Song2025Nature,matsuda2025multiferroiccollinearantiferromagnethidden}.
Therefore, understanding the properties of the relativistic p-wave Rashba- and Weyl- spin-orbit effect in combination with altermagnetism is crucial for further distinguishing between the properties of the two effects.
Our results, initially developed for the testbed material, can be extended to more general cases, as demonstrated by our analytical models. Furthermore, our research provides an extended and comprehensive analysis of various possible scenarios in altermagnets, particularly those involving the breaking of structural symmetries under relativistic effects, thereby offering deeper insights into their fundamental properties.


\begin{acknowledgments}
A. F., G. C., and C. A. contributed equally to this work.
C. A. acknowledges M. Cuoco, F. Forte, A.M. Le\'on and J.W. González for useful discussions.
This research was supported by the Foundation for Polish Science project “MagTop” no. FENG.02.01-IP.05-0028/23 co-financed by the European Union from the funds of Priority 2 of the European Funds for a Smart Economy Program 2021–2027 (FENG). G.C., S.P. and C. A. acknowledge support from PNRR MUR project PE0000023-NQSTI. We further acknowledge access to the computing facilities of the Interdisciplinary Center of Modeling at the University of Warsaw, Grant g91-1418, g91-1419, g96-1808, g96-1809 and g103-2540 for the availability of high-performance computing resources and support. We acknowledge the access to the computing facilities of the Poznan Supercomputing and Networking Center, Grants No. pl0267-01, pl0365-01 and pl0471-01.
\end{acknowledgments}

\bibliography{references}
\end{document}